\documentclass[a4paper,oneside]{article}
\usepackage[T1]{fontenc}
\usepackage{syntonly}
\usepackage[ansinew]{inputenc}

\usepackage{tocbibind}
\usepackage{amsmath}
\usepackage{amssymb}
\usepackage{bm}
\usepackage{physics}
\usepackage{multirow}
\usepackage{subcaption}
\usepackage{threeparttable}
\usepackage{longtable}
\usepackage{animate}
\usepackage{cite}
\usepackage{authblk}
\usepackage{morefloats}
\usepackage{subcaption}
\usepackage{caption}
\makeatletter
\newcommand*{\centerfloat}{%
  \parindent \z@
  \leftskip \z@ \@plus 1fil \@minus \textwidth
  \rightskip\leftskip
  \parfillskip \z@skip}
\makeatother
\DeclareCaptionLabelSeparator{bar}{~\rule[-0.4ex]{0.2ex}{1em}~}
\DeclareCaptionLabelFormat{subfor}{\textbf{#2}}
\usepackage{bibentry}
\usepackage[pdftex]{color,graphicx}
\usepackage[update,prepend]{epstopdf}

\author[1,2]{Ricard Alert\thanks{ricard.alert@princeton.edu}}
\affil[1]{Princeton Center for Theoretical Science, Princeton University, Princeton, NJ (USA)}
\affil[2]{Lewis-Sigler Institute for Integrative Genomics, Princeton University, Princeton, NJ (USA)}
\author[3,4]{Jaume Casademunt\thanks{jaume.casademunt@ub.edu}}
\affil[3]{Departament de F\'{i}sica de la Mat\`{e}ria Condensada, Universitat de Barcelona, Barcelona (Spain)}
\affil[4]{Universitat de Barcelona Institute of Complex Systems (UBICS), Barcelona (Spain)}
\author[5,6]{Jean-Fran\c{c}ois Joanny\thanks{jean-francois.joanny@college-de-france.fr}}
\affil[5]{Coll\`{e}ge de France, Paris (France)}
\affil[6]{Laboratoire PhysicoChimie Curie, Institut Curie, PSL University, Sorbonne Universit\'{e}s, UPMC, Paris (France)}

\title{Active Turbulence}

\date{\today}
\usepackage{xcolor}
\definecolor{UBcolor}{HTML}{007CC1}
\usepackage[colorlinks=true,pdfnewwindow=true,linkcolor=UBcolor,citecolor=UBcolor,urlcolor=UBcolor,breaklinks=true,linktocpage]{hyperref}
\usepackage[all]{hypcap}
\usepackage[capitalise,nameinlink]{cleveref}
\begin{document}

\maketitle

\begin{abstract}

Active fluids exhibit spontaneous flows with complex spatiotemporal structure, which have been observed in bacterial suspensions, sperm cells, cytoskeletal suspensions, self-propelled colloids, and cell tissues. Despite occurring in the absence of inertia, chaotic active flows are reminiscent of inertial turbulence, and hence they are known as \emph{active turbulence}. Here, we survey the field, providing a unified perspective over different classes of active turbulence. To this end, we divide our review in sections for systems with either polar or nematic order, and with or without momentum conservation (wet/dry). Comparing to inertial turbulence, we highlight the emergence of power-law scaling with either universal or non-universal exponents. We also contrast scenarios for the transition from steady to chaotic flows, and we discuss the absence of energy cascades. We link this feature to both the existence of intrinsic length scales and the self-organized nature of energy injection in active turbulence, which are fundamental differences with inertial turbulence. We close by outlining the emerging picture, remaining challenges, and future directions.

\end{abstract}

\clearpage

\tableofcontents

\section{Introduction} \label{intro}

Turbulence is a ubiquitous phenomenon. It takes place in systems ranging from stars and interstellar gas clouds to the atmosphere and the oceans, and down to the scales of our daily activities such as breathing, cooking, and driving. Turbulence is central in many scientific problems such as mixing processes and drag reduction, and also in engineering applications. In physics, turbulence is at the crossroads of fluid dynamics and statistical physics, challenging us to grasp their deep connections \cite{Frisch1995,Rose1978,Falkovich2006}.

Inertial turbulence is controlled by the Reynolds number, which embodies the ratio between inertial and viscous forces. At high Reynolds number, inertial effects destabilize laminar flow and lead to the chaotic patterns of vortices and jets that we know as turbulence. In a simplified picture of fully-developed inertial turbulence, the external driving injects kinetic energy at a certain scale, at which viscous dissipation is negligible. Due to inertial effects, this energy is then transported across scales through a so-called energy cascade, until it is dissipated by viscous effects at a very different scale. In the intermediate scales, known as the inertial range, the flow exhibits self-similarity, and velocity correlations are scale-invariant --- a property that is usually studied via the energy spectrum of the flow. In seminal work in 1941, Kolmogorov predicted that the energy spectrum follows a power law as a function of wave number, $E(q)\sim q^{-5/3}$, with a universal exponent independent of both the external driving and the fluid's properties. This picture of turbulence, involving the emergence of universal statistical properties of the flow from the combination of driving and nonlinearity, has become a paradigm of non-equilibrium physics \cite{Kolmogorov1991,Frisch1995,Rose1978,Kraichnan1980,Falkovich2006,Boffetta2012}.

Two decades ago, seemingly chaotic, multiscale flows were discovered in polymer solutions at low Reynolds numbers, for which inertial effects are negligible. In this case, nonlinear elastic effects due to the polymer destabilize the laminar flow and lead to so-called \emph{elastic turbulence} \cite{Groisman2000}. This type of turbulence is controlled by either the Deborah or the Weissenberg numbers, which compare elastic and viscous effects. The flow exhibits scale-invariant properties parallel to those of inertial turbulence albeit with different universal exponents \cite{Groisman2001,Steinberg2019}.

A few years after the discovery of elastic turbulence, spontaneous chaotic flows at low Reynolds numbers were found in a bacterial suspension \cite{Dombrowski2004}. They were initially called \emph{bacterial turbulence}. In the following decade, turbulent-like flows were discovered and characterized in several other active fluids, often of biological origin (\cref{Fig1}). Examples include different bacterial suspensions \cite{Dombrowski2004,Cisneros2007,Sokolov2007,Ishikawa2011,Sokolov2012,Wensink2012,Dunkel2013,Patteson2018,Li2019,Peng2020,Liu2020}, swarming sperm \cite{Creppy2015}, suspensions of microtubules and molecular motors \cite{Sanchez2012,Henkin2014,Guillamat2017,Lemma2019,Martinez-Prat2019,Tan2019,Martinez-Prat2021}, tissue cell monolayers \cite{Doostmohammadi2015,Yang2016a,Blanch-Mercader2018,Lin2021}, and suspensions of artificial self-propelled particles \cite{Nishiguchi2015,Karani2019}.

The defining feature of active matter is that it is driven internally by its constituents, which transform stored free energy into motion \cite{Marchetti2013}. Active systems are prone to experience instabilities and self-organization phenomena, thus developing correlated collective flows at large scales. When these self-driven flows become spatiotemporally chaotic, the situation is analogous to inertial turbulence at a descriptive level. Hence, these flows are commonly known as \emph{active turbulence} \cite{Thampi2016a}, but also as mesoscale turbulence or simply spatiotemporal chaos. Here, we use the name active turbulence. Beyond this descriptive analogy, there has been recent progress in establishing to what extent and in which sense active chaotic flows define new classes of turbulence. The question, however, remains unsettled and requires further work.

A key difference between active and inertial turbulence is worth emphasizing from the start. In inertial turbulence, the scale at which energy is injected is imposed by the external driving. In contrast, active flows are autonomous, and the pattern of energy injection is self-organized. In other words, the spectrum of energy injection is not an input of the problem but part of its solution. In active matter, the energy input is ultimately due to its microscopic components, be them  self-propelled colloids, swimming bacteria or molecular motors. However, their motion is correlated over much larger scales. Therefore, in a continuum description of the system as a fluid, the energy input into the hydrodynamic modes occurs at the characteristic scales of the flows. Consequently, both energy injection and dissipation are peaked at similar scales, and hence one may not expect energy cascades spanning an arbitrarily large range of scales as in traditional turbulence. Nevertheless, the spectra of energy injection and dissipation need not be exactly equal, thus allowing for energy transfer across scales. Finally, regardless of energy cascades, active turbulence can exhibit scale-free correlations. As we will discuss, the scaling regimes of active turbulence showcase similarities and differences with inertial turbulence. 

Here, we review the different types of active turbulence, in an effort to provide a unified perspective. A key challenge is that, in contrast to the Navier-Stokes equation for inertial turbulence, in active turbulence there is no unique fundamental equation but rather a variety of equations to describe diverse situations. In terms of approach, there are two classes of equations: (i) phenomenological generalizations of the Navier-Stokes equation, adding terms that inject energy and produce instabilities, and (ii) the hydrodynamic equations for active liquid crystals, which derive from
symmetries and conservation laws.
Moreover, active turbulence usually takes place in systems with orientational order. Turbulence is qualitatively different in systems with either polar or nematic order, and we have thus organized the review according to this distinction.
In each section, we also distinguish between `wet' and `dry' systems. Following the convention in the active matter literature, dry systems
do not conserve momentum, usually due to the presence of a frictional substrate or environment, whereas wet systems are dominated by hydrodynamic interactions and are described by momentum-conserving equations. This distinction is not inherent to the physical system but to the model; a system may be described as wet or dry depending on parameter values and
observation scales. Before discussing theory, we briefly describe experiments.

\section{Experiments} \label{experiments}

\begin{figure*}[tbp]
\vskip-2.5cm
\begin{center}
\includegraphics[width=0.8\textwidth]{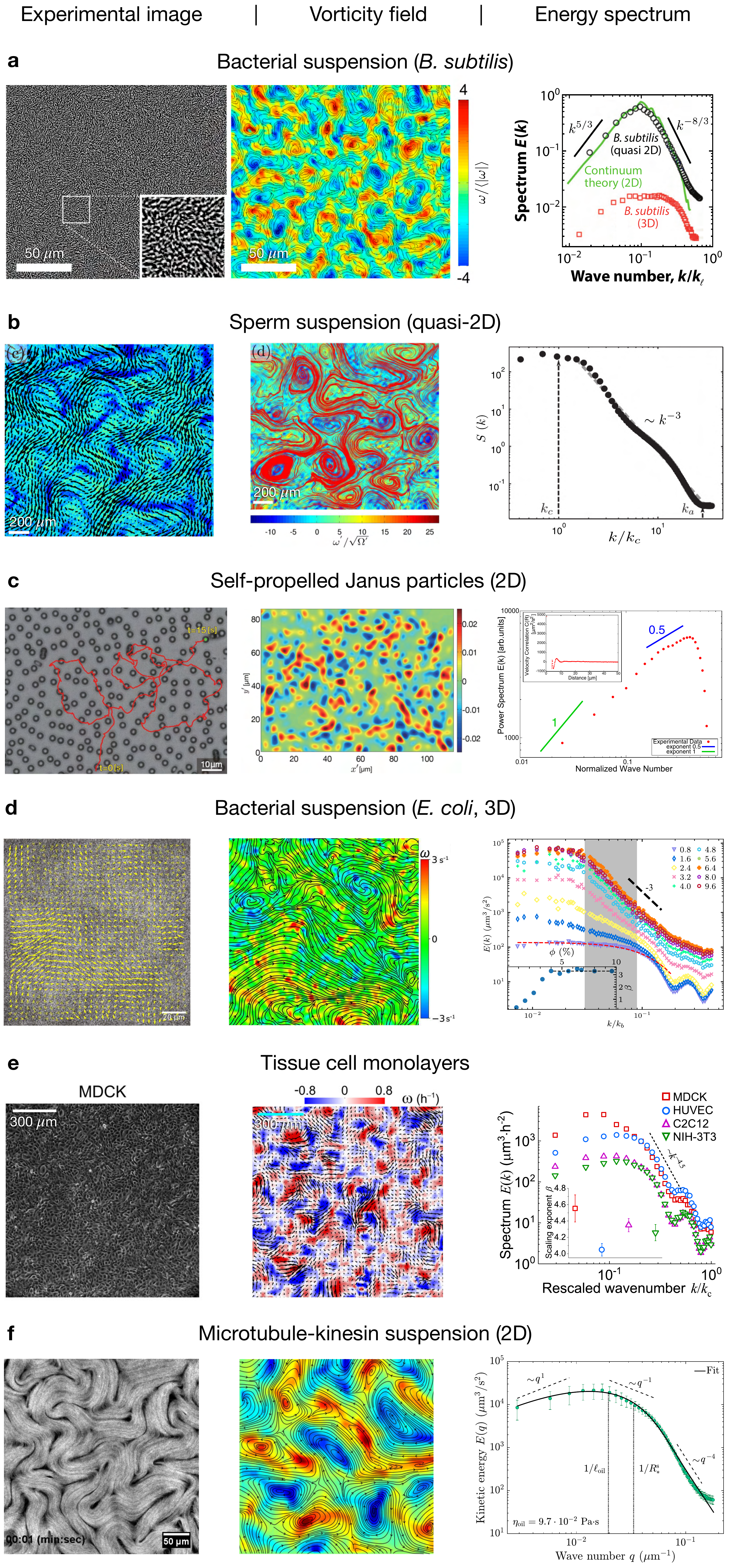}
  {\phantomsubcaption\label{Fig bacteria-2D}}
  {\phantomsubcaption\label{Fig sperm}}
  {\phantomsubcaption\label{Fig Janus}}
  {\phantomsubcaption\label{Fig bacteria-3D}}
  {\phantomsubcaption\label{Fig tissue}}
  {\phantomsubcaption\label{Fig microtubules}}
\caption{\textbf{Experiments on active turbulence}. For each experimental system, the three columns show an image of the system, a snapshot of the vorticity field (with either streamlines or the flow field in some cases), and the measured energy spectrum $E(k)\propto k \langle|\tilde{\bm{v}}_{\bm{k}}|^2\rangle$. \subref*{Fig bacteria-2D}, Adapted from \cite{Wensink2012}. \subref*{Fig sperm}, Adapted from \cite{Creppy2015}. \subref*{Fig Janus}, Adapted from \cite{Nishiguchi2015}. The inset shows the velocity correlation function from which the spectrum is obtained. \subref*{Fig bacteria-3D}, Adapted from \cite{Peng2020,Liu2020}. The different spectra correspond to different volume fractions of bacteria. The inset shows the scaling exponent of the spectrum as a function of the volume fraction. \subref*{Fig tissue}, Adapted from \cite{Lin2021}. The different spectra correspond to different cell types. \subref*{Fig microtubules}, Adapted from \cite{Guillamat2017,Martinez-Prat2021}.}
\label{Fig1}
\end{center}
\end{figure*}

Since its original discovery in bacterial suspensions, active turbulence has been found in an increasing number of systems (\cref{Fig1}). Some are based on active particles. Microswimmers such as bacteria (\cref{Fig bacteria-2D,Fig bacteria-3D}, sperm cells (\cref{Fig sperm}), and self-propelled colloids (\cref{Fig Janus}) are polar entities with head-tail asymmetry. Often, these particles align via both collisions and hydrodynamic interactions, leading to active polar phases that exhibit collective motion, equivalent to the flocking of birds. However, polar swimmers can potentially also align in a head-tail-symmetric way, leading to nematic phases. Determining the type of cell alignment in dense bacterial suspensions remains challenging. Other systems are clearly nematic, as revealed by the presence of half-integer topological defects. For example, in dense monolayers, eukaryotic cells have anisotropic shapes and align along a common axis (\cref{Fig tissue}). In these conditions, cells lack a persistent polarity, and therefore align nematically. Similarly, under the action of a depletant agent, polar cytoskeletal filaments known as microtubules align nematically into bundles (\cref{Fig microtubules}). The bundles are then actively sheared by kinesin molecular motors that crosslink and walk on parallel filaments, producing active nematic forces.

In addition to their varied symmetries, these systems also lie at different points along the wet-dry spectrum. For example, bacteria in dense suspensions are thought to interact mainly by steric repulsion and other short-range interactions such as lubrication forces \cite{Drescher2011,Koch2011}. Hence, dense bacterial suspensions have been successfully modeled as dry systems (\cref{Fig bacteria-2D} and \cref{dry-polar}). Yet, long-range hydrodynamic interactions might dominate in more dilute suspensions, where bacterial turbulence has been modeled in the wet limit (\cref{Fig bacteria-3D,wet-polar,wet-nematics}). Respectively, cohesive cell monolayers lie on solid substrates, with which they can exchange momentum by friction. Hence, they are often treated as dry systems. Yet, cell monolayers behave as wet systems at length scales smaller than the hydrodynamic screening length $\lambda = \sqrt{\eta/\xi}$, where $\eta$ is the monolayer viscosity and $\xi$ is the cell-substrate friction coefficient. This length typically ranges from a few tens to several hundreds microns depending on the strength of cell-cell and cell-substrate adhesion \cite{Blanch-Mercader2017,Duclos2018,Perez-Gonzalez2019}; it can therefore encompass from a few cells up to a few hundreds, placing the system in the wet-dry crossover (\cref{dry-nematics}). Finally, two-dimensional microtubule films can be assembled at an oil-water interface. Hence, there is also a viscous screening length $\ell_{\text{v}} = \eta/\eta_{\text{ext}}$ that results from comparing the two-dimensional viscosity $\eta$ of the active nematic film with the three-dimensional viscosity $\eta_{\text{ext}}$ of the external passive fluids. This system therefore also exhibits a wet-dry crossover \cite{Martinez-Prat2021} (\cref{Fig microtubules,dry-nematics}).

Over the years, experiments have characterized some statistical properties of active turbulence. For example, motivated by theoretical predictions, experiments on active nematics have measured the vortex area distribution \cite{Guillamat2017,Lemma2019,Martinez-Prat2021,Blanch-Mercader2018}. The central quantity common to all types of active turbulence is the velocity correlation function or, equivalently, the kinetic energy spectrum, which has been measured in most experimental systems \cite{Wensink2012,Liu2020,Creppy2015,Martinez-Prat2021,Lin2021,Nishiguchi2015} (\cref{Fig1}). Although these results are very informative, further measurements are required to fully characterize the scaling regimes and sort systems into classes. Moreover, additional relevant quantities have not been measured yet. These quantities include the spectra of energy transfer across scales, as well as the spectra of elastic energy in nematics. In addition to revealing this crucial information, future research might also discover active turbulence in other systems such as vibrated granular rods \cite{Narayan2007}, actin-based liquid crystals \cite{Kumar2018,Zhang2018c}, cilia-driven flows on bronchial tissues \cite{Loiseau2020}, and nematic bacterial colonies migrating on substrates \cite{Copenhagen2021}.

\section{Polar fluids} \label{polar}

The observation of turbulent-like flows in bacterial suspensions motivated the first theories of active turbulence. Early approaches were based on models of microswimmer suspensions, which lead to continuum equations for three coupled fields: swimmer concentration, coarse-grained polarity, and fluid velocity \cite{Aranson2007,Wolgemuth2008}. Simulations of these equations produced chaotic flows similar to the experiments.

Later work modelled bacterial turbulence ignoring the solvent flow, i.e. treating the system as dry. For example, simulations of self-propelled particles, either rods with steric interactions \cite{Wensink2012,Bar2020} or points with competing alignment interactions \cite{Grossmann2014}, yield a number of phases including irregular vortex patterns similar to the experimental observations (\cref{Fig SPR}). At the continuum level, Wensink, Dunkel, et al. proposed a phenomenological description based on a single equation for the velocity field of the bacterial suspension \cite{Wensink2012,Dunkel2013,Dunkel2013a}. This equation, known as the Toner-Tu-Swift-Hohenberg (TTSH) equation, has been the main focus of research in dry polar active turbulence \cite{Wensink2012,Dunkel2013,Dunkel2013a,Bratanov2015,James2018a,James2018}. We discuss its origin and main results in \cref{TTSH,TTSH-scaling}. More recently, the TTSH equation has been derived by explicit coarse-graining from microscopic models of microswimmers \cite{Heidenreich2016,Reinken2018}. Thus, this work connects the initial wet suspension model with the final dry description in terms of a single velocity field. Finally, in \cref{active-polar-LC} we discuss other types of turbulence in models of active polar liquid crystals with friction.

Work in wet polar systems has mainly been in two directions. One follows the phenomenological approach of the TTSH equation to propose a momentum-conserving extension of the Navier-Stokes equation. We discuss it in \cref{GNS}. The other direction involves further work on active polar suspensions, which uncovered the mechanisms of the transition to spatiotemporal chaos in these systems. We discuss it in \cref{active-polar-suspensions}.

\subsection{Dry systems} \label{dry-polar}

\subsubsection{Toner-Tu-Swift-Hohenberg equation} \label{TTSH}

The TTSH equation builds on the Toner-Tu equations for flocking \cite{Toner1995,Toner2005,Marchetti2013}, which provide a coarse-grained description of the Vicsek model for self-propelled aligning particles \cite{Vicsek1995,Marchetti2013}. As in flocking, the TTSH equation describes the coarse-grained velocity $\bm{v}$ of the particles, and it ignores hydrodynamic interactions mediated by the solvent. The solvent, however, exchanges momentum with the particles. Thus, the momentum of the particle system is not conserved, making the theory dry. This description does not distinguish the velocity from the polarity field and assumes that motion always occurs along the particle orientation direction. Finally, in contrast to the original Toner-Tu equations, the TTSH assumes incompressibility, i.e. $\bm{\nabla}\cdot\bm{v} = 0$.

\begin{figure*}[tb]
\begin{center}
\includegraphics[width=\textwidth]{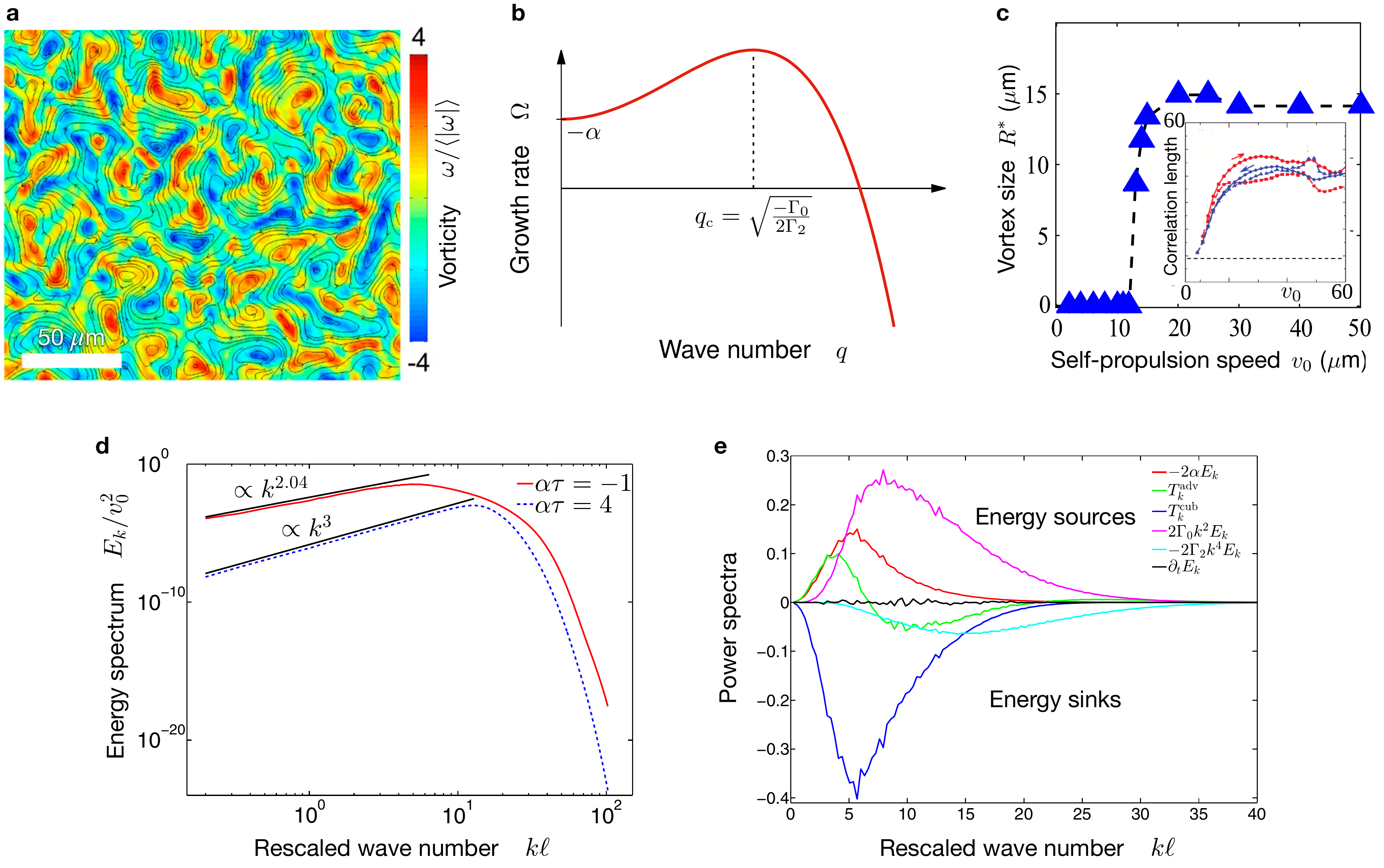}
  {\phantomsubcaption\label{Fig SPR}}
  {\phantomsubcaption\label{Fig growth-rate-TTSH}}
  {\phantomsubcaption\label{Fig vortex-size-swimmers}}
  {\phantomsubcaption\label{Fig spectrum-TTSH}}
  {\phantomsubcaption\label{Fig transfer-TTSH}}
\caption{\textbf{Active turbulence in dry polar fluids}. \subref*{Fig SPR}, Snapshot of the vorticity field and streamlines obtained from a simulation of self-propelled rods, which produces results similar to those of the continuum TTSH equation. Adapted from \cite{Wensink2012}. \subref*{Fig growth-rate-TTSH}, Growth rate of velocity perturbations in the TTSH equation for $\alpha<0$, $\Gamma_0<0$, and $\Gamma_2>0$. In this regime, the quiescent state is unstable at long wavelengths, producing vortices with a characteristic size $\sim q_{\text{c}}^{-1}$. \subref*{Fig vortex-size-swimmers}, Simulations of the TTSH equation derived from a microscopic model of self-propelled swimmers also show the selection of a characteristic vortex size that saturates at high activity. A similar trend is found for the velocity correlation length measured experimentally in 3D bacterial suspensions \cite{Sokolov2012} (inset). Adapted from \cite{Heidenreich2016}. \subref*{Fig spectrum-TTSH}, The energy spectrum of the TTSH equation exhibits a scaling regime with a non-universal exponent that varies with the parameter $\alpha$. This parameter controls energy injection, from lower (blue) to higher (red). \subref*{Fig transfer-TTSH}, Spectra of the contributions to the energy balance in the stationary turbulence regime of the TTSH equation ($\alpha\tau=-1$). The advective nonlinearity transfers energy across scales. Panels \subref*{Fig spectrum-TTSH} and \subref*{Fig transfer-TTSH} adapted from \cite{Bratanov2015}.}
\label{Fig2}
\end{center}
\end{figure*}

Following the structure of the Toner-Tu equation, the TTSH equation can be expressed as \cite{Wensink2012}
\begin{equation} \label{eq Toner-Tu}
\left( \partial_t + \bm{v}\cdot \bm{\nabla} \right) \bm{v} = -\bm{\nabla} P \ -(\alpha + \beta |\bm{v}|^2 ) \bm{v} + \bm{\nabla} \cdot \bm{E}
\end{equation}
where the pressure $P$ is determined by the incompressibility constraint $\bm{\nabla}\cdot \bm{v} =0$.
The Landau-like expansion in powers of the velocity describes the flocking transition. For $\alpha<0$, with $\beta>0$, the system develops long-range polar order, flowing with velocity $v_0=\sqrt{-\alpha/\beta}$ in an arbitrary direction. Respectively, the strain-rate tensor $\bm{E}$ takes the form
\begin{equation}
E_{i j} = S \left( v_i v_j - \frac{1}{d} |\bm{v}|^2 \delta_{i j} \right) + \left( \Gamma_0 - \Gamma_2 \nabla^2 \right) \left( \partial_i v_j + \partial_j v_i \right). 
\end{equation}
The first term, with $d$ the space dimension, corresponds to anisotropic active nematic stresses due to the microswimmers. These stresses give contributions to the advection and the pressure terms, and \cref{eq Toner-Tu} can be rewritten as \cite{Wensink2012}
\begin{equation} \label{eq TTSH}
\left( \partial_t + \lambda_0 \bm{v}\cdot \bm{\nabla} \right) \bm{v}  + \bm{\nabla} \left( p -\lambda_1|\bm{v}|^2 \right) =
\left[  \Gamma_0 \nabla^2 - \Gamma_2 \nabla^4 -(\alpha + \beta |\bm{v}|^2) \right] \bm{v},
\end{equation}
where $\lambda_0=1-S$ and $\lambda_1=-S/d$. Note that, because of the absence of Galilean invariance, the coefficient $\lambda_0$ of the self-advection term can be different from one. For puller swimmers, such as the algae \emph{C. reinhardtii}, $S>0$. For pushers, such as the bacterium \emph{B. subtilis}, $S<0$. The $\Gamma_0$ term corresponds to effective viscous stresses in the suspension.

All the terms discussed so far are taken from the original Toner-Tu equation. However, to destabilize the flocking state and produce vortices, the TTSH equation assumes the
effective viscosity to be negative, $\Gamma_0<0$ \cite{Wensink2012}. Even though this is a phenomenological assumption, negative effective viscosities are possible in active suspensions due to a combination of extensile activity and flow alignment, which decrease and may perhaps even reverse the viscous resistance to shear \cite{Hatwalne2004,Sokolov2009,Giomi2010,Ramaswamy2010,Marchetti2013,Gachelin2013,Lopez2015,Saintillan2018}. Moreover, derivations of the TTSH equation from microscopic models reveal that activity indeed leads to a decreased effective viscosity $\Gamma_0$, which could possibly turn negative \cite{Grossmann2014,Heidenreich2016,Reinken2018}. Note that there are important active or non-equilibrium contributions in three terms: the negative friction $\alpha$, the self-advection term $\lambda_0$, and the negative viscosity $\Gamma_0$.

With $\Gamma_0<0$, one needs the next-order term in the velocity gradient expansion, with $\Gamma_2>0$, to stabilize the flow at short wavelengths. The right-hand side of \cref{eq TTSH} has the
structure of the Swift-Hohenberg equation for pattern formation, originally obtained in the context of thermal convection \cite{Swift1977,Cross1993}. The right-hand side of \cref{eq TTSH} produces a long-wavelength instability, which leads to energy injection across the range of unstable modes.
The growth rate has a maximum at a finite wave number, $q_c = \sqrt{-\Gamma_0/(2\Gamma_2)}$, which defines a characteristic vortex size (\cref{Fig growth-rate-TTSH}). This is consistent with both microswimmer models \cite{Heidenreich2016} and experiments \cite{Sokolov2012}, which reveal a characteristic velocity correlation length that saturates at high activity (\cref{Fig vortex-size-swimmers}). $\Gamma_0$ and $\Gamma_2$ also define a characteristic velocity $V_\Gamma = \sqrt{|\Gamma_0|^3/\Gamma_2}$. The existence of these characteristic scales of velocity and vortex size, at which the energy spectrum is maximal (\cref{Fig bacteria-2D}), is an important difference with inertial turbulence, which features scale-free vortex distributions.

Overall, the quiescent, isotropic state ($\bm{v}=0$) of the TTSH equation is subject to two activity-driven instabilities, which respectively yield a uniform transition to flocking due to a negative friction ($\alpha<0$), and spatial pattern formation due to a negative viscosity ($\Gamma_0<0$). The combination of these two instabilities with the nonlinear effects of velocity self-advection lead to turbulent states in the TTSH equation. In fact, velocity self-advection makes the TTSH equation non-potential, whereas the right-hand side can be derived from an effective free energy and would therefore not produce chaos.
The parameters in \cref{eq TTSH} can be fitted to achieve good quantitative agreement with the experimentally-measured velocity correlation functions and energy spectra (\cref{Fig bacteria-2D}), both in 2D \cite{Wensink2012} and in 3D \cite{Dunkel2013} dense bacterial suspensions. These spectra exhibit scaling regimes described by power laws, which we discuss next.

\subsubsection{Power laws with non-universal exponents} \label{TTSH-scaling}

At scales larger and smaller than the typical vortex size, the spectrum obtained in Ref. \cite{Wensink2012} exhibits scaling regimes with exponents close to $5/3$ and $-8/3$, respectively (\cref{Fig bacteria-2D}). With the simplifying choice $\lambda_1=0$ in \cref{eq TTSH}, later work analyzed TTSH turbulence in more detail, searching for parallels and differences with inertial turbulence \cite{Bratanov2015}. This study revealed that the scaling exponents are non-universal, in the sense that they depend on parameter values (\cref{Fig spectrum-TTSH}).
Non-universal behavior in the TTSH equation arises from its cubic nonlinearity ($\beta |\bm{v}|^2\bm{v}$), which is not present in the Navier-Stokes equation. Combined with the advective nonlinearity ($\bm{v}\cdot\bm{\nabla}\bm{v}$) common to inertial turbulence, the cubic Toner-Tu term provides additional freedom, allowing the system to self-organize into turbulent states with parameter-dependent statistical properties.
In this sense, the authors of Ref. \cite{Bratanov2015} claim that these properties define a new class of turbulence.

The spectral energy balance reveals that energy injection is indeed controlled by the $\alpha$ and $\Gamma_0$ terms driving the instabilities \cite{Bratanov2015}. The $\Gamma_2$ term contributes to dissipation. As in 2D Navier-Stokes turbulence, the advective nonlinearity transports energy from intermediate to large scales. The additional cubic nonlinearity dominates the dissipation at large scales (\cref{Fig transfer-TTSH}). In particular, it dissipates the energy transported by the advective effects, thus preventing the inverse energy cascade of inertial 2D turbulence \cite{Bratanov2015}. More recent work has confirmed this scenario and revealed that a very strong advective transfer leads to the emergence of vortex patterns from a transient turbulent state \cite{James2018a}.

\subsubsection{Active polar liquid crystals with friction} \label{active-polar-LC}

Other types of dry turbulence can take place in models for active polar liquid crystals with friction. In contrast to the TTSH equation, and like in two-phase suspensions, in liquid crystals the velocity and the polarity fields can be different from one another. In fact, polarity-velocity misalignment occurs in systems such as Janus colloids \cite{Zhang2020} and epithelial monolayers \cite{Kim2013,Notbohm2016,Alert2020}, in which cells polarize in response to mechanical stresses and environmental cues \cite{Alert2020}. Similarly, early experiments and simulations on bacterial turbulence found a low correlation between the velocity and polarity fields, suggesting a strong advection of bacteria by the solvent \cite{Sokolov2007,Aranson2007}.

Despite being simpler than full microswimmer suspensions, turbulence in polar liquid crystals has not received as much attention. An exception is Ref. \cite{Blanch-Mercader2017c}. Motivated by mechanical waves in epithelial cell monolayers on solid substrates, this study proposed an active polar fluid model based on the following force balance: $\bm{\nabla}\cdot\bm{\sigma} = \xi\bm{v} - T_0 \bm{p}$. Here, $\bm{\sigma}$ is the stress tensor of the fluid, which includes viscous, active, and flow alignment contributions. The interaction forces with the substrate include viscous friction with coefficient $\xi$ and active traction with magnitude $T_0$ in the direction of cell polarity $\bm{p}$. In turn, the polarity evolves as dictated by liquid crystal hydrodynamics \cite{DeGennes-Prost,Marchetti2013,Prost2015,Julicher2018}, which includes advection and co-rotation of the polarity by the flow.

The polar state, with $|\bm{p}|=1$ and speed $v_0 = T_0/\xi$, is destabilized by the active stresses. Moreover, the advection term $\bm{v}\cdot\bm{\nabla}\bm{p}$ in the polarity dynamics endows the growth rate with an imaginary part, which leads to nonlinear traveling waves. Near the instability threshold, modulations of these waves follow the so-called complex Ginzburg-Landau equation, originally proposed in the context of pattern formation \cite{Aranson2002,Cross1993}. This equation exhibits several forms of spatiotemporal chaos, including phase turbulence, amplitude turbulence, spatiotemporal intermittency, and bistable chaos \cite{Chate1994}. These states could therefore arise via secondary bifurcations from the nonlinear waves \cite{Blanch-Mercader2017c}, as also suggested in Ref. \cite{Marcq2014}. Whether such forms of active turbulence are found in polarized epithelial tissues or other polar systems remains an open question.

\subsection{Wet systems} \label{wet-polar}

\subsubsection{Generalized Navier-Stokes equation} \label{GNS}

Following the phenomenological approach of the TTSH equation, S\l{}omka and Dunkel proposed the so-called Generalized Navier-Stokes (GNS) equation to describe active turbulence in polar momentum-conserving systems \cite{Slomka2015}. In contrast to the TTSH equation, the GNS equation does not include Toner-Tu flocking terms. Therefore, it contains neither the negative friction leading to polar order nor the cubic nonlinearity that enables non-universal scaling in the TTSH equation (see \cref{TTSH,TTSH-scaling}). Instead, the activity is phenomenologically encoded in a postulated, generic extension of the stress tensor. This approach is again similar in spirit to the Swift-Hohenberg equation for pattern formation \cite{Swift1977,Cross1993}, and it had also been used in the context of granular media \cite{Aranson2006}. Here, the effective stress tensor is proposed to be linear in the velocity field, and it is expressed as a gradient expansion, thus conserving momentum. The only nonlinearity of the GNS equation is thus the velocity self-advection term, as in the Navier-Stokes equation. Assuming incompressibility ($\bm{\nabla}\cdot\bm{v} = 0$), the GNS equation reads
\begin{equation} \label{eq GNS}
\left( \partial_t + \bm{v}\cdot \bm{\nabla} \right) \bm{v} = -\bm{\nabla}P  + \bm{\nabla} \cdot \bm{\sigma},
\end{equation}
with
\begin{equation} \label{eq eff_visc}
\sigma_{i j} = \left[ \Gamma_0 - \Gamma_2 \nabla^2 + \Gamma_4 \nabla^4 \right] \left( \partial_i v_j + \partial_j v_i \right).
\end{equation}
Except for the $\Gamma_2$ and $\Gamma_4$ terms, this equation has the form of the Navier-Stokes equation. In contrast to the TTSH equation, here $\Gamma_0$ is a positive viscosity. Therefore, we may think of this model as having an effective kinematic viscosity that depends on the spatial scale, which is represented in Fourier space via the following dependence on the wave number $q$: $\tilde{\nu}_{\text{eff}} (q) = \Gamma_0 +\Gamma_2 q^2 + \Gamma_4 q^4$. Turbulence at low Reynolds number is then modeled as arising from an instability obtained by setting $\Gamma_2<0$ while keeping $\Gamma_4>0$. This choice produces a band of unstable modes which, unlike for the TTSH equation (\cref{Fig growth-rate-TTSH}), exclude the longest wavelengths ($q\rightarrow 0$). The unstable modes inject energy to drive flows and, as in the TTSH equation, they produce patterns of vortices with a characteristic size. These patterns can then become turbulent thanks to the nonlinear advection term inherited from the Navier-Stokes equation \cite{Slomka2015}.

The GNS equation was proposed as an alternative model for the bacterial turbulence experiments. As for the dry TTSH equation, the parameters of the GNS equation were also fitted to the velocity correlation functions and probability distributions of 3D bacterial suspensions \cite{Slomka2017}, as originally measured in Refs. \cite{Wensink2012,Dunkel2013}. On the theoretical side, the relative mathematical simplicity of the GNS equation enabled analytical solutions of vortex lattices \cite{Slomka2017a}. It was also formulated in a covariant way to simulate active turbulence on curved surfaces \cite{Mickelin2018}. In 3D, the theory predicted an inverse energy cascade \cite{Slomka2017,Slomka2018}, opposite to the direct cascade of inertial 3D turbulence. The active inverse cascade is also different from the inverse cascade of inertial 2D turbulence in that it is not driven by vortex merging. Instead, the active cascade involves the formation of moving vortex chains, and it results from a spontaneous breaking of chiral symmetry \cite{Slomka2017,Slomka2018}. Observing these effects in experiments remains a challenge.

Finally, a variant of the GNS equation was also used to study the crossover from active to inertial 2D turbulence as the Reynolds number is increased. By varying the effective viscosity spectrum $\tilde{\nu}_{\text{eff}}(q)$, simulations revealed a sharp transition between the small vortices characteristic of GNS active turbulence and the emergence of a classic inverse cascade that accumulates energy at the largest scales, giving rise to a so-called condensate of two large vortices \cite{Linkmann2019,Linkmann2020}. The authors argue that this transition might be observable in bacterial suspensions, which could reach moderate Reynolds numbers owing to the reduction of their effective viscosity by activity (see discussion in \cref{TTSH}). The active-to-inertial transition might also be observable in other systems where turbulence is driven by larger active objects such as magnetic spinners \cite{Kokot2017} and camphor boats \cite{Bourgoin2020}, which might reach even higher Reynolds numbers.

\begin{figure*}[tb]
\begin{center}
\includegraphics[width=0.9\textwidth]{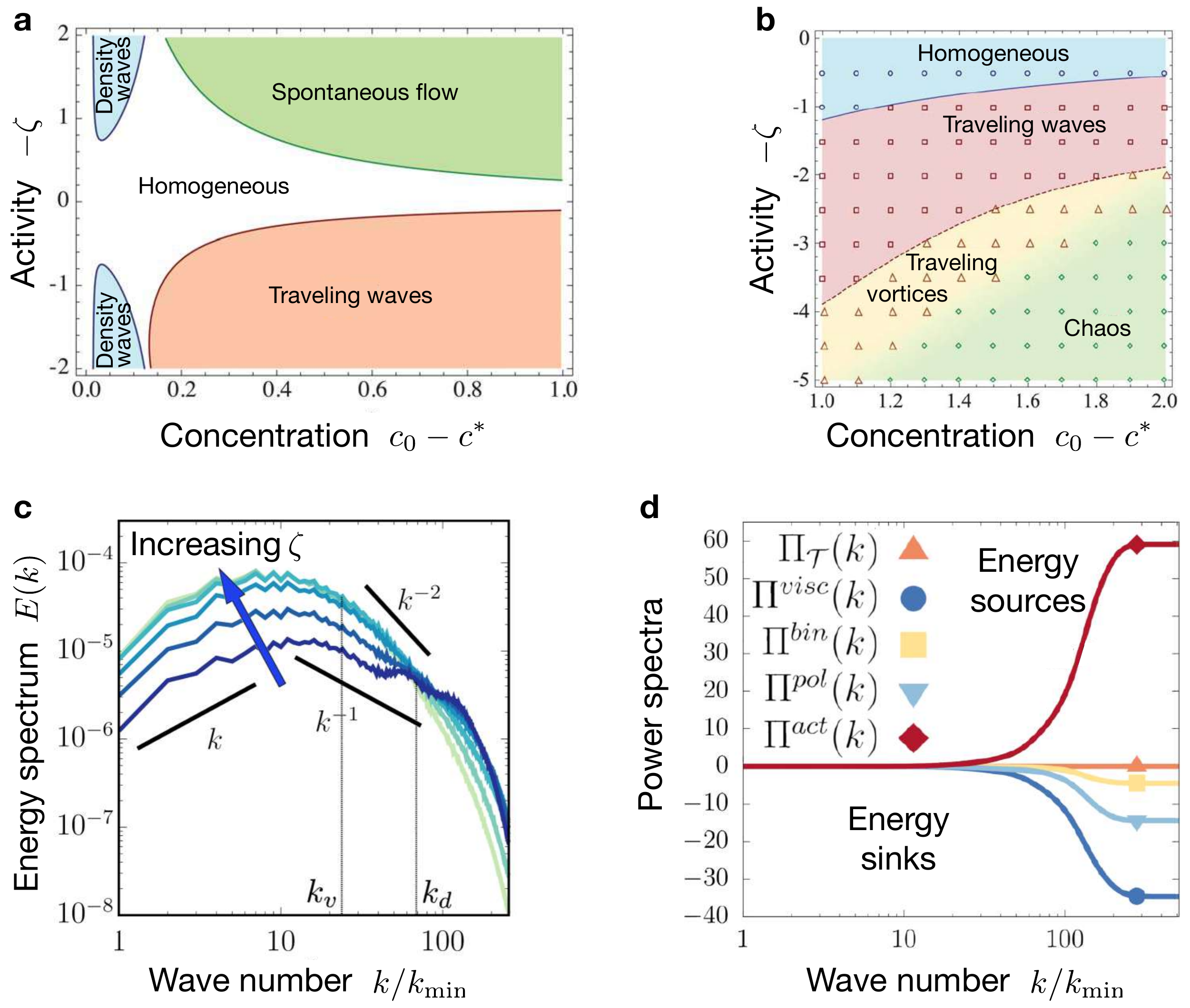}
  {\phantomsubcaption\label{Fig stability-polar-suspensions}}
  {\phantomsubcaption\label{Fig patterns-polar-suspensions}}
  {\phantomsubcaption\label{Fig spectrum-polar-suspensions}}
  {\phantomsubcaption\label{Fig transfer-polar-suspensions}}
\caption{\textbf{Transition to chaos and active turbulence in wet polar suspensions}. \subref*{Fig stability-polar-suspensions}, Phase diagram obtained from a linear stability analysis of the homogeneous state. Density waves are like those found in the Toner-Tu equations, i.e. in the dry limit. In addition to spontaneous flow, wet polar suspensions can exhibit traveling waves. \subref*{Fig patterns-polar-suspensions}, Nonlinear spatiotemporal patterns obtained in simulations of the continuum equations, revealing the emergence of chaos via traveling waves and vortices. Panels \subref*{Fig stability-polar-suspensions} and \subref*{Fig patterns-polar-suspensions} adapted from \cite{Giomi2012}. \subref*{Fig spectrum-polar-suspensions}, The energy spectrum in simulations of wet polar emulsions features activity-dependent scaling regimes. $k_{\text{v}}$ is the first moment of the energy distribution, and $k_{\text{d}}$ is related to the scale of emulsion droplets. \subref*{Fig transfer-polar-suspensions}, Spectra of the contributions to the energy balance in simulations of active polar emulsions, which exhibit no energy transfer across scales (orange). The other contributions are associated with viscosity (dark blue), the binary composition of the system (yellow), passive polar stresses (light blue), and activity (red). Panels \subref*{Fig spectrum-polar-suspensions} and \subref*{Fig transfer-polar-suspensions} adapted from \cite{Carenza2020}.}
\label{Fig3}
\end{center}
\end{figure*}

\subsubsection{Active polar liquid crystals and suspensions} \label{active-polar-suspensions}

As in the dry case, wet polar systems can also be modeled using the formalism of liquid crystal hydrodynamics. In contrast to the phenomenological approach of the GNS equation (\cref{GNS}), the hydrodynamic equations of polar liquid crystals are established by symmetries and conservation laws \cite{DeGennes-Prost,Marchetti2013,Prost2015,Julicher2018}, and the parameters are measurable quantities with a direct physical interpretation. The hydrodynamic equations are similar to those of active nematics (see \cref{nematic-hydrodynamics}), replacing the director field by a polarity field and adding polar terms that break the invariance under $\bm{p}\rightarrow -\bm{p}$.

Work on wet turbulence in one-component polar liquid crystals has been surprisingly scarce \cite{Bonelli2016}. Most work has focused on two-component suspensions, whose equations include a swimmer concentration field $c$ \cite{Aranson2007,Wolgemuth2008,Giomi2008,Tjhung2011,Giomi2012}. In this case, active stresses become proportional to the concentration field. Moreover, the swimmers self-propel with respect to the fluid, and therefore the concentration and the polarity fields are advected both by the fluid and by the swimmers. Accordingly, the material derivatives in the transport equations read $\partial_t c + \bm{\nabla}\cdot \left[ (\bm{v} + w_c \bm{p}) c\right]$ and $\left[\partial_t + (\bm{v} + w_p \bm{p})\cdot \bm{\nabla}\right] \bm{p}$. Here, $w_c$ and $w_p$ characterize the self-advection terms arising from self-propulsion, which were found to enhance concentration inhomogeneities \cite{Tjhung2011}.

Polarity self-advection is a feature common with polar dry systems; it appears in the Toner-Tu equations. Consequently, active polar suspensions share some of the instabilities and spatiotemporal patterns of dry polar fluids \cite{Giomi2012}, such as the traveling density waves found in the Toner-Tu equations \cite{Tu1998,Bertin2009,Geyer2018,Marchetti2013} (\cref{Fig stability-polar-suspensions}). Moreover, the hydrodynamic equations of polar fluids reduce to those of nematic fluids when the polar terms are ignored. Hence, active polar suspensions share the spontaneous flow instability of active nematics (\cref{Fig stability-polar-suspensions}), which we discuss in \cref{spontaneous-flow}. Yet other instabilities arise from the interplay between advection and active stresses, leading to hydrodynamic traveling waves \cite{Giomi2012} (\cref{Fig stability-polar-suspensions}) reminiscent of experimental observations in bacterial swarms \cite{Chen2017a}. These waves, as those discussed in \cref{active-polar-LC}, are specific to polar fluids.

Unlike in nematic fluids, instabilities in polar fluids can occur via Hopf bifurcations, giving rise to oscillatory phenomena. Simulations of polar suspensions indeed find oscillatory flows that become increasingly complex, with the sequential addition of frequencies in the spectrum, as activity increases \cite{Giomi2008}. In fact, more complete simulations revealed a route to spatiotemporal chaos through patterns of traveling waves and vortices \cite{Giomi2012} (\cref{Fig patterns-polar-suspensions}). Simulations of polar liquid crystals with polarity-velocity alignment painted a similar picture, with oscillatory and traveling patterns leading to chaos \cite{Bonelli2016}. Intriguingly, this sequence of patterns was also observed in simulations of active nematics confined between two walls with different friction \cite{Ramaswamy2016}. Apart from this specific case, the route to chaos through Hopf bifurcations is different than that of nematic systems, either the one-component fluids that we discuss in \cref{wet-nematics}, or suspensions, which exhibit excitable behavior prior to the transition to chaos \cite{Giomi2011}.

The chaotic flow regime is only beginning to be characterized. Recent work by Carenza et al. has obtained the statistical properties of turbulent-like flows in simulations of active polar emulsions, in which the active component is localized in the emulsion droplets \cite{Carenza2020}. The results show energy spectra featuring three scaling regimes, and the exponent of the intermediate regime seems to vary with activity (\cref{Fig spectrum-polar-suspensions}). Predicting these results remains an open challenge. Furthermore, the energy injected by active stresses is dissipated by a combination of passive effects (\cref{Fig transfer-polar-suspensions}). Hence, the chaotic flows involve no energy transfer across scales, and therefore no turbulent cascades.

Finally, recent work by \v{S}kult\'{e}ty et al. has used kinetic theory to coarse-grain a microscopic model of microswimmers with hydrodynamic interactions. Going beyond the usual mean-field approximation, this work showed that, below the onset of collective motion, velocity correlations are suppressed by particle self-propulsion \cite{Skultety2020}.

\section{Nematic fluids} \label{nematic}

The early observations of bacterial turbulence led to the theories for active polar turbulence that we discussed in \cref{polar}. In this section, we discuss theories of active nematic turbulence, which were motivated by chaotic flows reported first in microtubule suspensions (\cref{Fig microtubules}) and then in epithelial monolayers (\cref{Fig tissue}). Like for polar fluids, both dry and wet models have been put forward, as well as models that focus on the wet-dry crossover. Unlike in \cref{polar}, however, we discuss wet systems first and dry systems later. Before, we give a general introduction to active nematic hydrodynamics, discussing the spontaneous-flow instability that leads to active nematic turbulence, and the dimensionless number that controls it.

\subsection{Active nematics} \label{nematic-general}

\subsubsection{Hydrodynamic equations} \label{nematic-hydrodynamics}

Most theoretical work on turbulence in active nematics is based on hydrodynamic equations for macroscopic slow variables \cite{Ramaswamy2010,Marchetti2013,Prost2015,Julicher2018}. These 
variables are either conserved quantities, such as concentrations and momentum, or soft 
modes arising from the spontaneous breaking of continuous symmetries such as isotropy.

In nematics, orientational order is described by a traceless symmetric tensor known as the nematic order parameter, $\bm{Q}$. It can be expressed as $\bm{Q} =\frac{2}{3} S \bm{q}$, where $q_{ij}=n_{i}n_{j}-\frac{1}{d}\delta_{ij}$ and $d$ is the space dimension \cite{DeGennes-Prost}. Here, $\hat{\bm{n}}$ is the so-called director field, which indicates the axis of alignment. Its direction is a soft mode. Respectively, $S$ is the amplitude of the order parameter, which quantifies the degree of alignment. Strictly speaking, it is not a soft mode. However, it can be useful to keep it in hydrodynamic equations to study situations in which the nematic order varies sharply, such as around topological defects. This approach is followed in most numerical work on active nematic turbulence, which we describe in the sections below.

The free energy $\mathcal{F}$ of a passive nematic is
a functional of the order parameter. The free energy per unit volume $F$ is 
split into two terms \cite{DeGennes-Prost,Doostmohammadi2018,Thampi2016a}: the Maier-Saupe free energy $F_{\text{MS}}$ describes the isotropic-nematic transition and it depends only on the
order parameter $\bm{Q}$; the Frank free $F_{\text{F}}$ energy describes the orientational elasticity of the nematic and it depends on the gradients of $\bm{Q}$.

Deep in the nematic phase, the amplitude of the order parameter is $S=1$, and 
it is sufficient to describe the nematic order by the director orientation, given by its angle $\theta$
in two dimensions. The free energy then reduces to
the Frank contribution. In the one-constant approximation \cite{DeGennes-Prost}, it reads $F_{\text{F}} = \frac{K}{2}(\bm{\nabla}\bm{Q})^2$, where $K$ is the Frank elastic constant. The relaxation of this energy is driven by the so-called orientational field $\bm{h}=
-\frac{\delta \mathcal{F}}{\delta \bm{n}}$. In two dimensions, its 
component perpendicular to $\bm{n}$ is $h_{\perp}=K \nabla^2 \theta$ \cite{DeGennes-Prost}, which tends to make the director relax to a uniform configuration.

In addition to relaxing according to the orientational field, the nematic order parameter is also advected by the flow $\bm{v}$, co-rotated by the vorticity $\bm{\omega} = \bm{\nabla}\times\bm{v}$, and it realigns in response to shear, which is described by the symmetric shear-rate tensor $u_{ij} = \frac{1}{2}(\partial_i v_j + \partial_j v_i)$. Taking all these effects into account, the dynamics of the director field is given by
\begin{equation} \label{eq director}
\partial_t \bm{n} + \bm{v}\cdot \bm{\nabla}\bm{n} + \frac{1}{2}\bm{\omega}\times\bm{n} = \frac{\bm{h}}{\gamma} - \nu \bm{u}\cdot\bm{n}.
\end{equation}
Here, $\gamma$ is the rotational viscosity, and $\nu$ is the flow-alignment parameter. This parameter depends on the shape of the nematic particles: $\nu<0$ for rod-like or prolate particles, and $\nu>0$ for disk-like or oblate particles. If $|\nu|>1$, the director tends to align at a fixed angle with respect to the shear. In contrast, for $|\nu|<1$, bulk passive nematics exhibit the so-called tumbling instability whereby the director rotates without settling at any particular angle. \Cref{eq director} specifies the director dynamics when nematic order is fully described by the director field, and for an incompressible nematic ($\bm{\nabla}\cdot\bm{v}=0$). The structure of the equation 
for the full order parameter is very similar, see
Refs. \cite{Doostmohammadi2018,Thampi2016a}. The main difference is that all the transport coefficients 
defined above depend on the amplitude $S$ of the order parameter.

The hydrodynamic theory of passive nematics has been generalized to active nematics \cite{Kruse2005,Marchetti2013,Prost2015,Julicher2018}. In this case, the stress tensor can be decomposed into active and passive contributions: $\bm{\sigma}=\bm{\sigma}_{\text{pas}} + \bm{\sigma}_{\text{act}}$. The passive stress 
includes the pressure $P$, three elastic contributions, and the viscous stress: ${\bm 
\sigma}_{\text{pas}}=-P {\bm I}+{\bm \sigma}_{\text{E}} +{\bm \sigma}_{\text{a}} +{\bm \sigma}_{\nu} +{\bm \sigma}_{\text{v}}$. 
The Ericksen stress $\bm{\sigma}_{\text{E}}$ is a 
non-isotropic contribution to the pressure; in two dimensions, it is given by 
$\sigma_{\text{E}}^{ij}=K \partial_{i}\theta \partial_{j}\theta 
-\frac{K}{2}|\bm{\nabla} \theta|^2\delta_{ij}$. The antisymmetric stress $\sigma_{\text{a}}^{ij}=\frac{1}{2} 
\left(n_{j}h_{j}-n_{j}h_{i}\right)$ arises from torques associated with the conservation of angular momentum. The third elastic term is the conjugate to the flow-alignment term in 
\cref{eq director}: $\sigma_{\nu}^{\alpha\beta}=\frac{\nu}{2} 
\left(n_{i}h_{j}+n_{j}h_{i}\right)$. An 
incompressible nematic has three independent viscosities that we assume here equal and given by $\eta$.
The viscous stress is then ${\bm \sigma}_{\text{v}} =2\eta \bm{u}$ , where $\bm{u}$ is the symmetric shear-rate tensor defined above. Finally, the active stress is proportional to the orientational tensor: ${\bm 
\sigma}_{\text{act}}=-\zeta \bm{Q}$. It is extensile along the direction of the director 
for $\zeta>0$, and it is contractile for $\zeta<0$.

These constitutive equations must then be completed by the conservation laws. Here, we focus on systems at low Reynolds numbers, for which inertia is 
negligible. Force balance then reads 
\begin{equation}
    \partial_{i} \sigma^{ij}=0.
    \label{eq stress}
\end{equation}
For dry systems, in contact with a substrate or environment, the right-hand side of \cref{eq stress} would include additional forces such as friction or active traction, as we discuss in \cref{dry-nematics}.

\subsubsection{Spontaneous-flow instability} \label{spontaneous-flow}

One of the general properties of active nematics
is that the quiescent steady 
state with uniform director field is unstable, leading to spontaneous flows \cite{Voituriez2005,Simha2002,Edwards2009}.
The growth rate of orientational perturbations around the uniform state, with a wave vector $\bm{q}$ at an angle $\phi$ from the director, is given by \cite{Duclos2018}
\begin{equation}
    \Omega(q)= \frac{\zeta \cos2\phi (1-\nu \cos2\phi)}{2\left(\eta +\gamma\nu^2 \sin^2 2\phi/4\right)} -\frac{Kq^2}{\gamma} \frac{\eta + \gamma/4(\nu^2-2\nu \cos2\phi+1)}{\eta +\gamma\nu^2 \sin^2 2\phi/4}.
\end{equation}
This growth rate turns positive at long wavelengths (\cref{Fig growth-rate-nematic}), in a way that depends on the angle $\phi$ between the director and the wave vector. For $\phi=0$, the perturbations are along the direction of alignment, corresponding to bend distortions of the director field (\cref{Fig instability-mechanism} left). In this case, the quiescent state is 
unstable if $\zeta(1-\nu) \geq 0$. Thus, for rod-like objects ($\nu<0$), the instability to bend distortions occurs for extensile active stresses ($\zeta>0$, \cref{Fig instability-mechanism} left). For $\phi=\pi/2$, the perturbations are perpendicular to the direction of alignment, corresponding to splay distortions of the director field (\cref{Fig instability-mechanism} right). In this case, the instability occurs
if $\zeta (\nu+1)\leq 0$. Thus, for rod-like objects, splay distortions produce an instability if active stresses are contractile ($\zeta<0$, \cref{Fig instability-mechanism} right).
In all cases, the instability is based on the following mechanism: a director perturbation produces active forces, given by $\bm{\nabla}\cdot\bm{\sigma}_{\text{act}} = -\zeta\bm{\nabla}\cdot(\bm{n}\bm{n})$, which drive flows that further distort the director, amplifying the original perturbation (\cref{Fig instability-mechanism}). Director distortions offer less resistance for longer wavelengths, and hence the maximal growth rate is found at $q\rightarrow 0$ (\cref{Fig growth-rate-nematic}). As a result, right past the instability threshold, the pattern of spontaneous flows has a wavelength 
given by the system size.

Overall, active nematics are generically unstable at long wavelengths. The critical wave number $q_{\text{c}} = \ell^{-1}_{\text{a}} f(\nu, \eta/\gamma,\phi)$ is mainly set by so-called active length $\ell_{\text{a}}=\sqrt{K/|\zeta|}$, above which active stresses overcome the restoring elastic stresses of the nematic.  
As activity increases, the active length decreases, and hence the band of unstable modes becomes wider (\cref{Fig growth-rate-nematic}).

\subsubsection{Dimensionless numbers and the high-activity limit} \label{dimensionless}

Finally, the hydrodynamic equations of active nematics that we have introduced can be made dimensionless by rescaling lengths by the system size $L$, time by the active time $\tau_{\text{a}} = \eta/|\zeta|$ (\cref{Fig growth-rate-nematic}), and stresses by the active stress coefficient $|\zeta|$. The equations then contain only three dimensionless parameters: 
the viscosity ratio $\gamma/\eta$, the flow-alignment parameter $\nu$,
and the activity number $A=\left(L/\ell_{\text{a}}\right)^2$.

Active turbulence occurs in the limit of high activity number, corresponding to large 
systems and/or large active stresses. In this limit, among the three elastic contributions to the 
stress (see \cref{nematic-hydrodynamics}), only the flow-alignment term is not negligible with respect to the active stress at length scales 
larger than $\ell_{\text{a}}$. This is because all stresses proportional to the Frank constant $K$ (including via $h_\perp$) scale like $1/A$, and only $h_\parallel$ yields non-vanishing contributions at high activity. To obtain $h_\parallel$, we project \cref{eq director} on the director, which gives $h_{\parallel} = \gamma \nu \ \bm{n} \cdot \bm{u} \cdot \bm{n}$. Using this result, we express the flow-alignment stress as ${\bm \sigma}_{\nu}=\gamma \nu^2 (\bm{n} \cdot \bm{u} \cdot 
\bm{n})(\bm{n}\bm{n}) $. This expression shows that this elastic stress is equivalent to a viscous stress with an anisotropic viscosity tensor. Therefore, we expect it not to modify the scaling properties of active nematic turbulence, which we discuss next.

\subsection{Wet systems} \label{wet-nematics}

\subsubsection{Transition to turbulence} \label{transition-turbulence}

\begin{figure*}[tb!]
\begin{center}
\centerfloat
\includegraphics[width=1.2\textwidth]{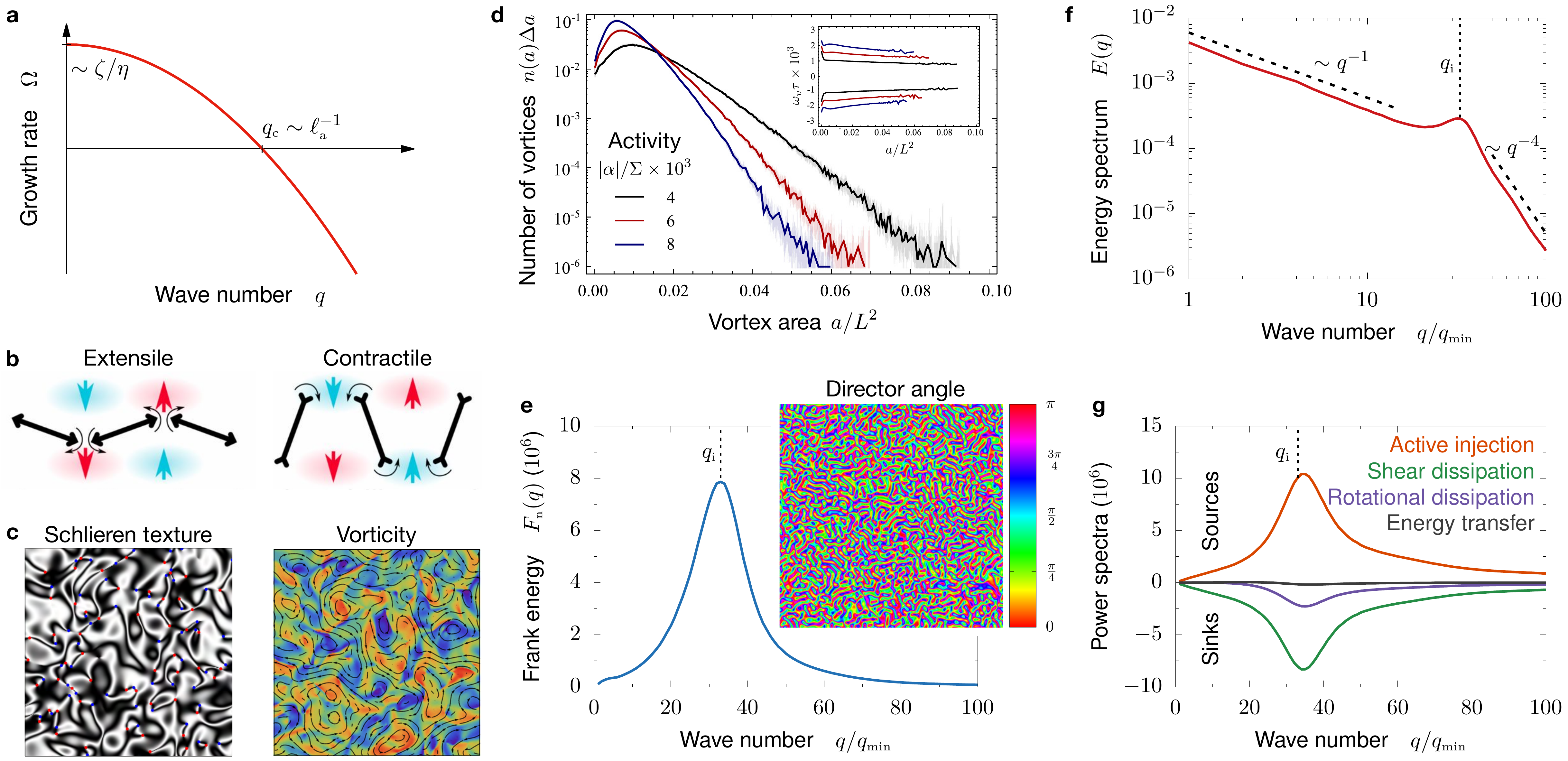}
  {\phantomsubcaption\label{Fig growth-rate-nematic}}
  {\phantomsubcaption\label{Fig instability-mechanism}}
  {\phantomsubcaption\label{Fig defects-vorticity}}
  {\phantomsubcaption\label{Fig vortex-distribution}}
  {\phantomsubcaption\label{Fig Frank-spectrum}}
  {\phantomsubcaption\label{Fig spectrum-nematic}}
  {\phantomsubcaption\label{Fig transfer-nematic}}
\caption{\textbf{Active turbulence in wet nematic fluids}. \subref*{Fig growth-rate-nematic}, Growth rate of director perturbations in wet active nematics. As discussed in the text, the system can exhibit a long-wavelength instability, with a critical wavelength proportional to the active length $\ell_{\text{a}}$. Increasing activity therefore widens the range of unstable modes. \subref*{Fig instability-mechanism}, Schematic of the instability mechanism. Flows generated by director perturbations further rotate the director field, amplifying the initial perturbation. Extensile systems can be unstable to longitudinal (bend) perturbations; contractile systems can be unstable to transverse (splay) perturbations. Adapted from \cite{Maitra2018}. \subref*{Fig defects-vorticity}, Snapshots of the director and the vorticity fields in simulations of active nematic turbulence. The director is represented by the Schlieren texture (see \cite{DeGennes-Prost}), with $+1/2$ and $-1/2$ defects indicated in red and blue, respectively. \subref*{Fig vortex-distribution}, Distribution of vortex areas in simulations show an exponential tail. The vorticity of individual vortices is rather independent of vortex size (inset). Panels \subref*{Fig defects-vorticity} and \subref*{Fig vortex-distribution} adapted from \cite{Giomi2015}. \subref*{Fig Frank-spectrum}, Spectrum of the Frank elastic energy of the director field in simulations of active nematic turbulence. The peak at $q_{\text{i}}$ is related to the characteristic size of the vortices and director domains shown in the inset. \subref*{Fig spectrum-nematic}, Energy spectrum in simulations of active nematic turbulence feature scaling regimes with universal exponents both above and below the characteristic vortex size. \subref*{Fig transfer-nematic}, Spectra of the contributions to the energy balance in simulations of a minimal model of wet active nematics. There is no energy transfer across scales; the energy injected at each scale is dissipated at that same scale by the combined effects of shear and rotational viscosity. Panels \subref*{Fig Frank-spectrum} to \subref*{Fig transfer-nematic} adapted from \cite{Alert2020a}.}
\label{Fig4}
\end{center}
\end{figure*}

Turbulence in active nematics ultimately emerges from the spontaneous-flow instability described in \cref{spontaneous-flow}. Right past the instability threshold, the instability leads to steady states with periodic flow patterns. As activity is increased, smaller-wavelength modes become unstable (\cref{Fig growth-rate-nematic}), which destabilizes the roughly-uniform regions in the flow patterns. Thus, starting from a uniform state, the system experiences a sequence of instabilities that lead to steady states with vortex patterns. Finally, at high activity, flows become unsteady, creating spatio-temporal chaos. This scenario for the transition to turbulence based on a sequence of instabilities was recently illustrated in simulations without topological defects \cite{Alert2020a}, and it is consistent with experimental observations on microtubule suspensions \cite{Martinez-Prat2019}.

Previous work had uncovered an overall similar picture, albeit enhanced by the presence and dynamics of topological defects, i.e. singular points of the nematic order \cite{kleman1989}. In two-dimensional nematics, the lowest-energy defects are those with topological charge $+1/2$ and $-1/2$. Because of their different structures, $+1/2$ defects are propelled by active forces whereas $-1/2$ defects are not \cite{Doostmohammadi2018,Aranson2019}. This effect facilitates the proliferation of defects in active nematics. When an active nematic spontaneously flows, it generates sharp walls with a steep gradient of the director field separating nematic domains. As the activity increases, the walls decay into pairs of $+1/2$ and $-1/2$ defects \cite{Thampi2014a,Thampi2014b,Thampi2016a}, which can unbind because of the self-propulsion of $+1/2$ defects. This process was first observed in experiments with active microtubule suspensions \cite{Sanchez2012}, and then studied numerically \cite{Giomi2013,Giomi2014a} and theoretically \cite{Shankar2018}. In recent years, the defect dynamics has been extensively characterized in several experimental realizations of active nematics \cite{Doostmohammadi2018}.

The emergence of defects locally destroys the nematic order, which can be restored by the annihilation of pairs of opposite-sign defects. At high activity, the continuous creation and annihilation of defect pairs drives the chaotic formation and destruction of nematic domains, leading to active turbulence. This process has therefore been analyzed in terms of the rates of creation and annihilation of defect pairs, whose balance gives the steady-state defect density \cite{Thampi2013,Thampi2014b,Giomi2015}.

The transition to turbulence in active nematics has also been studied in confined systems, for example in channels \cite{Fielding2011,Shendruk2017,Doostmohammadi2019,Doostmohammadi2017,Hardouin2019a}. As in unconfined systems, defect pairs are also generated above a critical activity. Under confinement, however, the defects organize in space and exhibit coordinated motions along an array of vortices rotating in alternating directions \cite{Shendruk2017,Doostmohammadi2019}, as recently observed in microtubule experiments \cite{Hardouin2019a}. Further increasing the activity, the final transition from vortex arrays to chaotic flows occurs through the intermittent appearance of locally disordered flow patches, known as active puffs. Below a critical activity, these puffs split and decay. Above a critical activity, however, the active puffs percolate through the channel, corresponding to the emergence of turbulence. As a function of the distance to the critical activity, the active-puff fraction grows as a power law, with an exponent that coincides with that of the direct percolation universality class \cite{Doostmohammadi2017}. Overall, the intermittent-like behavior and the connection to critical phenomena is reminiscent of similar phenomena in the transition to inertial turbulence in channels \cite{Frisch1995,Rose1978,Falkovich2006}. Thus, future work along these lines might reveal further connections between active and inertial turbulence.

\subsubsection{Scaling with universal exponents} \label{scaling-nematics}

Fully-developed active turbulence arises in 
the limit of high activity number. In two dimensions, the turbulent state 
appears as a disordered array of 
swirling regions, which are strongly coupled to defects of the nematic order (\cref{Fig defects-vorticity}). 
In this section, we focus on the structure and statistical properties of the flow and the director's orientation patterns. 

Giomi analyzed the turbulent regime in two-dimensional simulations (with parameters $\nu\ll 1$ and $\gamma\sim \eta$), focusing first on the statistical properties of 
the vorticity pattern \cite{Giomi2015}.
Vortices are identified as regions where the 
so-called Okubo-Weiss parameter, which measures the local Lyapunov exponent, is 
negative.
The statistics of the vortex areas $a$ is
well described by an exponential distribution $p(a)\sim \exp (a/a_*)$ (\cref{Fig vortex-distribution}), whose
characteristic vortex area $a_* = \pi R_*^2$ defines the average vortex radius $R_*$. This characteristic vortex size is proportional to the active length $\ell_{\text{a}}$, which is the only intrinsic length scale \cite{Hemingway2016a} (see \cref{spontaneous-flow,Fig growth-rate-nematic}).
Simulations also show that the individual vortices have a vorticity $\omega_0$ that is rather independent of their size (\cref{Fig vortex-distribution} inset), and which is inversely proportional to the active time (see \cref{dimensionless}): $\omega_0\sim \tau_{\text{a}}^{-1}\sim |\zeta|/\eta$.

Based on these observations, Giomi describes the
turbulent state by making two assumptions \cite{Giomi2015}: (i) the absolute value of the vorticity 
is uniform and equal to $\omega_0$ inside each vortex, and (ii) vortices are not 
correlated, which corresponds to a mean-field approximation. Averaging over the vortex area distribution, these two 
assumptions allow to calculate the spectrum of the enstrophy ${\Omega} =\int \frac{1}{2} 
\bm{\omega}^2(\bm{r})\,\dd\bm{r} =\int \frac{\dd q}{4\pi^2} \, {\tilde \Omega}(q)$, which characterizes the vorticity distribution in Fourier space. This calculation shows that the enstrophy spectrum $\tilde \Omega(q)$ features two scaling regimes. At scales smaller than the characteristic vortex size ($qR_*\gg 1$), it scales as
$\tilde \Omega(q)\sim q^{-2}$, and at large scales ($qR_*\ll 1$), it scales as 
$\tilde \Omega(q) \sim q$. Finally, the spectrum
of the kinetic energy 
per unit mass density $E= \int  \frac{1}{2} \bm{v}^2(\bm{r}) \,\dd\bm{r} = \int \frac{\dd q}{4\pi^2} \,\tilde E(q)$ is directly related to the enstrophy spectrum: $\tilde E(q)\sim \tilde \Omega(q)/q^2$. Therefore, the predicted scaling laws are
\begin{equation} \label{eq spectrum}
\begin{array}{ll}
\tilde E(q)\sim q^{-1} & \text{for}\; qR_*\ll 1,\\
\tilde E(q)\sim q^{-4} & \text{for}\; qR_*\gg 1.
\end{array}
\end{equation}
The scaling at small scales is well observed in Giomi's simulations \cite{Giomi2015}. 
However, the activity number was not large enough to observe the scaling at large scales.

More recently, Alert et al. have performed simulations in two dimensions (with $\nu=0$ and $\gamma \sim \eta$) at much higher activity number, reaching $A=3.5\times 10^5$ \cite{Alert2020a}. This study formulated a minimal theory that considers only the nematic orientation angle, assuming a fixed strength $S=1$ of the order parameter as we introduced in \cref{nematic-hydrodynamics}. Thus, this formulation does not allow for the spontaneous creation of defects. Yet, simulations show disordered patterns of elongated orientation domains that exhibit persistent dynamics that looks chaotic in both space and time (\cref{Fig Frank-spectrum} inset). The orientational pattern is statistically isotropic and has a characteristic wavelength proportional to the active length $\ell_{\text{a}}$, as revealed by the spectrum $\tilde F(q)$ of the elastic Frank free energy
${\mathcal F} =\int \frac{1}{2} |\bm{\nabla} \theta|^2 \dd \bm{r} = \int \frac{\dd q}{4\pi^2} \, \tilde F(q)$, which is peaked around a wave number $q_{\text{i}}\sim 1/\ell_{\text{a}}$ (\cref{Fig Frank-spectrum}). At larger scales ($q\rightarrow 0$), the spectrum of the director angle scales as $\langle |
\tilde \theta (q)|^2\rangle \sim q^0$, meaning that angle correlations decay over the active length $\ell_{\text{a}}$, consistent with the existence of orientational domains of this characteristic size.

Regarding the flow spectrum, the simulations by Alert et al. reached sufficiently high activity to clearly show the two scaling regimes of \cref{eq spectrum}, separated by a crossover at $q_{\text{i}}\sim 1/\ell_{\text{a}}$ \cite{Alert2020a} (\cref{Fig spectrum-nematic}). Complementary to the calculations leading to \cref{eq spectrum}, the large-scale scaling $\tilde E(q)\sim q^{-1}$ can be predicted in the high-activity limit via the following argument. At distances large compared to $\ell_{\text{a}}$, director correlations can be considered as short range. Hence, the only non-vanishing contribution to the elastic stress, $\bm{\sigma}_{\nu}$ (see \cref{dimensionless}), can be locally averaged over the director orientation, giving $\langle \bm{\sigma}_{\nu}\rangle = \gamma \nu^2 \langle(\bm{n} \cdot \bm{u} \cdot \bm{n})(\bm{n} \bm{n}) \rangle = 2\frac{\gamma \nu^2}{8} 
\bm{u}$.
The elastic stresses therefore renormalize the shear 
viscosity to $\eta_{\text{eff}}= \eta (1 + \frac{\gamma \nu^2}{8\eta})$.

The statistical properties of the flow can then be predicted from the equation for the vorticity, which can be derived by taking the curl of the force balance \cref{eq stress}. Using dimensionless variables, and in the high-activity limit, the vorticity equation reads
\begin{equation} \label{eq vorticity}
   \left(1 + \frac{\gamma \nu^2}{8\eta}\right) \nabla^2\omega  = \mathrm{sgn}(\zeta)\left[\frac{1}{2} \left[\partial_x^2 - \partial_y^2\right] \sin 2\theta - \partial_{xy}^2 \cos 2\theta \right].
\end{equation}
This is a Poisson equation with a source 
that depends only on the director's angle $\theta$. As argued above, at large scales, the angle can be considered considered as a random 
variable which is delta-correlated in space. However, the propagator of the Laplace operator in \cref{eq vorticity} is long-ranged (varying with distance as $\log r$ in two dimensions), reflecting the long-range nature of hydrodynamic interactions. As a result, short-range angle correlations can lead to long-range correlations of the flow field. Accordingly, using $\langle | \tilde\theta (q) |^2 \rangle \sim q^0$ in \cref{eq vorticity} in Fourier space leads to $\langle | \tilde\omega (q) |^2 \rangle \sim q^0$. Finally, this result leads to the scaling of \cref{eq spectrum}, $\tilde E(q)\sim q^{-1}$, which corresponds to a velocity correlation function $\langle \bm{v}(0)\bm{v}(r)\rangle \sim \frac{K|\zeta|}{\eta_{\text{eff}}^2} \log r$.

These results establish that active nematic turbulence exhibits a universal scaling regime at large scales. The scaling exponent is universal: it does not depend on
the sign of the active stress; it is observed in simulations for both 
contractile and extensile stresses. It is also independent of the 
flow-alignment parameter $\nu$ and of the viscosity ratio $\gamma/\eta$. Furthermore, the defect-free simulations also demonstrate that the scaling properties of active nematic turbulence are not sensitive to the existence of topological defects \cite{Alert2020a}. Finally, recent microtubule suspension experiments are consistent with the predicted scaling regimes \cite{Martinez-Prat2021} (\cref{Fig microtubules}).

So far, our discussion has been restricted to two-dimensional turbulence, which has been the primary focus of both experimental and theoretical work. However, some experimental systems are three-dimensional. In fact, the scaling arguments above can be extended to three dimensions, predicting $\tilde E(q)\sim q^0$ and $\tilde E(q)\sim q^{-3}$ at scales respectively larger and smaller than the typical vortex size. These scalings have also been predicted in the context of microswimmer suspensions. Below the onset of collective motion, B\'{a}rdfalvy et al. proposed to treat swimmers as hydrodynamically-interacting force dipoles \cite{Bardfalvy2019a}. Assuming that the dipoles undergo random reorientations, the theory indeed predicts a plateau of the energy spectrum at large scales, $\tilde E(q)\sim c\kappa^2$, where $c$ is the swimmer concentration and $\kappa$ is the strength of the force dipoles.

Finally, these scaling laws for 3D flows, $\tilde E(q)\sim q^0$ and $\tilde E(q)\sim q^{-3}$, have been recently found in bacterial suspensions experiments \cite{Liu2020} (\cref{Fig bacteria-3D}). At first sight, this finding is surprising because bacterial turbulence has been successfully modeled using theories for polar fluids (\cref{polar,Fig bacteria-2D}), including recent generalizations to account for density inhomogeneities \cite{Worlitzer2021}. Yet, these results suggest that, at least below the onset of collective motion, the spectra in 3D are well described by the hydrodynamics of active nematics (\cref{nematic,Fig bacteria-3D}). Understanding if and how polar and nematic theories capture different regimes of bacterial turbulence in a consistent way will require further work.

\subsubsection{Absence of energy cascades} \label{cascades}

A hallmark of inertial turbulence is the existence of an energy cascade. In three dimensions, the kinetic energy injected by the driving at large scales is transported across the inertial range of scales until it is dissipated by viscous effects at small scales. The cascade is therefore called \emph{direct}: from large to small scales \cite{Kolmogorov1991,Frisch1995,Rose1978,Falkovich2006}. In two dimensions, however, the injected energy is transferred to larger scales; the energy cascade is \emph{inverse} \cite{Rose1978,Kraichnan1980,Boffetta2012,Falkovich2006}.

The situation in active fluids is entirely different. Recent work has revealed that there is no energy cascade in active nematic turbulence \cite{Alert2020a,Urzay2017,Carenza2020b}. Alert et al. derived the spectral energy balance in an active nematic fluid, which has four contributions \cite{Alert2020a}:
\begin{equation} \label{eq energy-balance}
\partial_t \tilde F (q) = - \tilde D_{\text{s}} (q) - \tilde D_{\text{r}} (q) + \tilde I(q) + \tilde T(q).
\end{equation}
The left-hand side corresponds to the power spectrum of the Frank elastic energy. The right-hand side contains, in order, the power spectra of the shear viscous dissipation ($2\eta \bm{u}^2$), the rotational viscous dissipation ($\bm{h}^2/\gamma$), the injection of energy by the active stress ($\sigma^{ij}_{\text{act}} u_{ij}$), and the energy transfer between scales, whose integral over $q$ vanishes in the steady state due to energy conservation. The simulation results showed that energy is injected at all scales, but primarily at the characteristic scale $q_{\text{i}}\sim \ell_{\text{a}}$ of the vortex and orientation patterns (\cref{Fig transfer-nematic}). Moreover, the simulations also revealed that, in the absence of flow alignment ($\nu=0$), all the energy injected at a given scale is dissipated at that same scale. Therefore, there is no energy left to be transferred to other scales, and hence there is no energy cascade \cite{Alert2020a} (\cref{Fig transfer-nematic}). The numerical results by Urzay et al. and Carenza et al. are consistent with this conclusion \cite{Urzay2017,Carenza2020b}.

Beyond the numerical results, an integral expression for the energy transfer spectrum $T(q)$ can be derived from the director dynamics. The advection of the director field in \cref{eq director} gives rise to potential energy transfer across scales. However, symmetry arguments show that this contribution to the energy transfer vanishes identically for the statistically-isotropic turbulent state \cite{Alert2020a}. Besides advection of the director, flow-alignment and other elastic stresses yield additional contributions to the energy transfer, which may be significant at scales comparable to the active length $\ell_{\text{a}}$. Indeed, recent simulations by Carenza et al. have uncovered some energy transfer at intermediate scales due to elastic stresses \cite{Carenza2020b}. However, they did not find the scaling regime $\tilde E(q)\sim q^{-1}$ at large scales (see \cref{scaling-nematics,Fig spectrum-nematic}), possibly due to not reaching sufficiently high activity numbers.

In fact, we expect the spectral energy balance to be simpler in the high-activity limit and in the scaling regime. By the same arguments given in \cref{dimensionless}, rotational dissipation scales like $1/A$ and hence becomes negligible in front of shear dissipation and injection. Similarly, the elastic contributions to energy transfer also scale like $1/A$ and therefore also become negligible. Finally, as also argued in \cref{dimensionless,scaling-nematics}, the flow-alignment contribution amounts to a renormalization of the shear viscosity. Thus, the emerging picture at high activity is that, in the scaling range, there is a simple scale-by-scale balance between energy injection and the renormalized viscous dissipation. At intermediate scales of order $\sim \ell_{\text{a}}$, where most energy injection and dissipation takes place, there can be energy transfer due to both flow-alignment and elastic nonlinearities.
Testing these predictions, both in simulations and in experiments, remains an important task for future work.

\subsection{Dry systems and the wet-dry crossover} \label{dry-nematics}

When interacting with a substrate or an external medium, the active fluid's momentum is not conserved, and hence the system is classified as dry \cite{Marchetti2013,Bar2020,chate2020}. Active turbulence in dry nematics can occur in systems of particles or rods with nematic interactions and whose concentration can vary. Microscopic models of these systems can be coarse-grained to obtain hydrodynamic equations for the order parameter and the density fields. Numerical work on these equations has revealed a rich phase diagram that includes regimes of spatiotemporal chaos \cite{Ngo2014a,Putzig2016,Maryshev2019,Patelli2019a,Bar2020,chate2020}. The chaotic behavior arises from the unstable dynamics of nematic density bands. Thus, turbulence in these systems is associated with density segregation, and thus also with giant number fluctuations \cite{Ngo2014a}.

A different type of dry active turbulence occurs in incompressible nematic fluids. In this case, the theoretical description starts from the hydrodynamic theory that we presented in \cref{wet-nematics}, and extends the force balance \cref{eq stress} to add interaction forces between the fluid and a substrate. These interaction forces may have both passive and active contributions, corresponding respectively to friction and traction forces. Friction comes in different types, which affect active turbulence in several ways that we discuss below. Before, we briefly introduce the active forces, whose role in active turbulence has not been particularly studied.

The functional form of interfacial active forces is dictated by nematic symmetry: $\bm{f}^{\text{act}} = \zeta_1 \bm{n} \bm{\nabla}\cdot\bm{n} + \zeta_2 \bm{n}\cdot\bm{\nabla}\bm{n}$. If momentum were conserved, the coefficients $\zeta_1$ and $\zeta_2$ would have to be equal so that the active force results from the divergence of a symmetric stress, $\bm{f}^{\text{act}} = \zeta \bm{\nabla}\cdot(\bm{n}\bm{n})$, as in \cref{nematic-hydrodynamics}. However, if the fluid exchanges momentum with the substrate, $\zeta_1$ and $\zeta_2$ can be different from one another. Suprisingly, these active forces can stabilize nematic suspensions against the well-known spontaneous-flow instability of \cref{spontaneous-flow} \cite{Maitra2018}.

\subsubsection{Substrate friction} \label{substrate-friction}

In nematic phases, friction can be anisotropic, with different coefficients for the directions parallel and perpendicular to the director $\bm{n}$: $\bm{f}^{\text{pass}} = -\xi \bm{v} + \Delta\xi\, \bm{n} (\bm{n}\cdot\bm{v})$. In fact, friction anisotropy leads to asymmetries in the flow field around topological defects, which has been found to lead to cell accumulation at defects in monolayers of tissue cells \cite{Kawaguchi2017} and bacteria \cite{Copenhagen2021}. Anisotropic friction can also organize otherwise chaotic flows into lanes with alternating directions \cite{Thijssen2020}, which had previously been discovered in experiments \cite{Guillamat2016}.

\begin{figure*}[tb]
\begin{center}
\includegraphics[width=\textwidth]{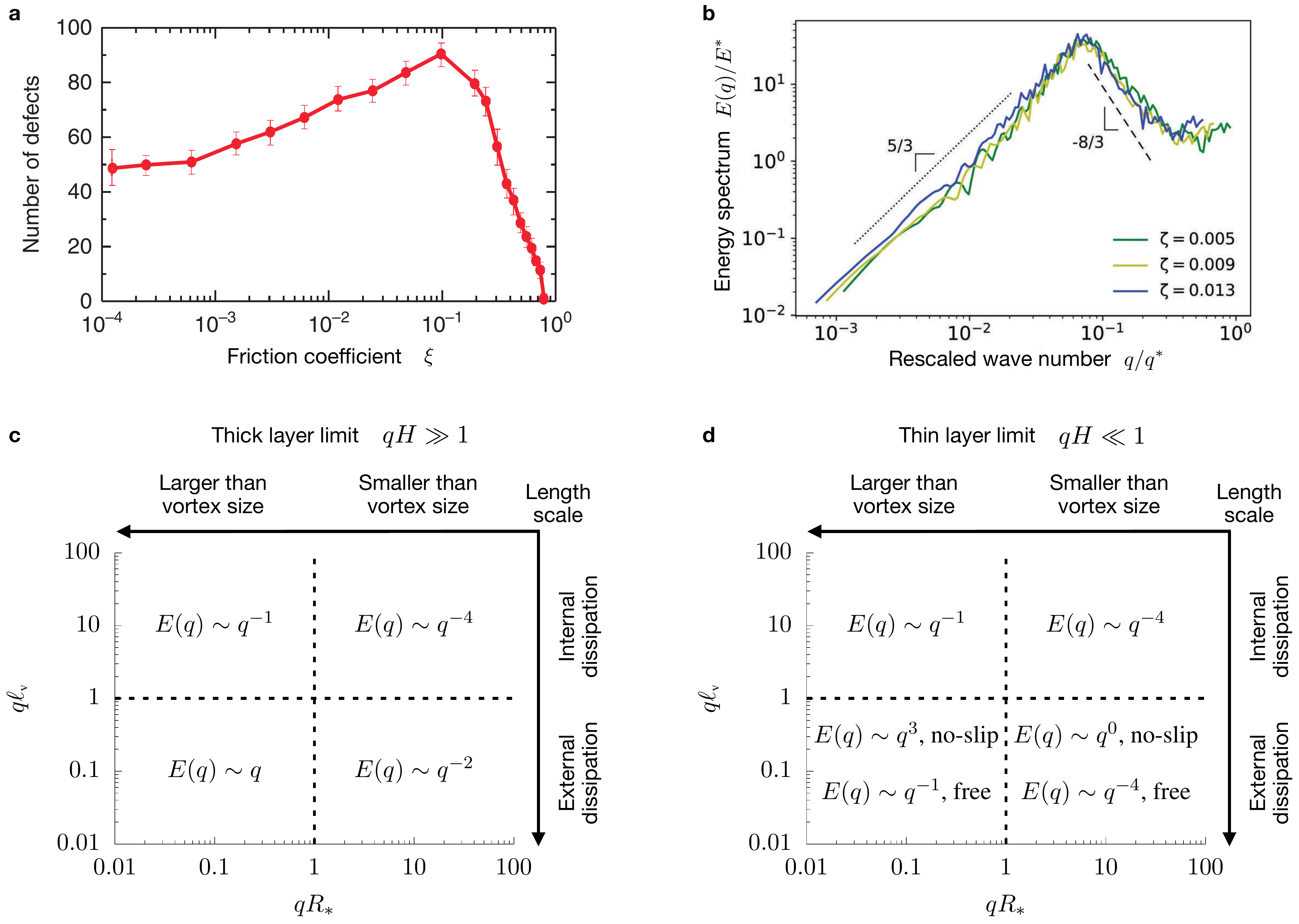}
  {\phantomsubcaption\label{Fig defect-number}}
  {\phantomsubcaption\label{Fig spectrum-dry-nematics}}
  {\phantomsubcaption\label{Fig scalings-thick-layer}}
  {\phantomsubcaption\label{Fig scalings-thin-layer}}
\caption{\textbf{Active turbulence in dry active nematics and the wet-dry crossover}. \subref*{Fig defect-number}, As friction increases, the number of defects in active nematic turbulence first increases and then decreases sharply as friction damps and stabilizes the flow. Adapted from \cite{Doostmohammadi2016}. \subref*{Fig spectrum-dry-nematics}, Energy spectrum in simulations of active nematic turbulence with friction. Adapted from \cite{Coelho2020}. These results show scaling regimes with exponents different than those predicted by theory. \subref*{Fig scalings-thick-layer},\subref*{Fig scalings-thin-layer}, Scaling regimes predicted for an active nematic film in contact with a passive fluid layer. This is a common experimental setup \cite{Martinez-Prat2021}. The scaling regimes are classified in terms of three lengths: the thickness of the passive fluid layer $H$ (panel \subref*{Fig scalings-thick-layer} vs \subref*{Fig scalings-thin-layer}), the average vortex size $R_*$, and the viscous length $\ell_{\text{v}} = \eta/\eta_{\text{ext}}$ above which dissipation is dominated by the external passive fluid. For thin external layers (\subref*{Fig scalings-thin-layer}), some of the scaling laws depend on whether the external fluid has a free surface or is in contact with a solid substrate (no-slip boundary condition). From \cite{Martinez-Prat2021}.}
\label{Fig5}
\end{center}
\end{figure*}

Even isotropic friction has a strong influence on active turbulence. Frictional dissipation dominates over viscous dissipation at length scales above the hydrodynamic screening length $\lambda = \sqrt{\eta/\xi}$, which therefore controls the crossover from wet to dry situations. This crossover has been studied in several simulations by adding friction to the hydrodynamic equations of active nematics (see \cref{nematic-hydrodynamics}) \cite{Thampi2014,Doostmohammadi2016,Srivastava2016a,Oza2016,Oza2016a,Doostmohammadi2019,Coelho2020,Thijssen2020b}. As friction increases and the screening length gets closer to the active length $\ell_{\text{a}}$ (see \cref{spontaneous-flow}), the number of defects actually first increases \cite{Thampi2014,Doostmohammadi2016} (\cref{Fig defect-number}). In these conditions, recent simulations have observed turbulent flows whose energy spectrum features scaling regimes with exponents close to $5/3$ and $-8/3$ \cite{Coelho2020} (\cref{Fig spectrum-dry-nematics}). These scaling laws, however, are different than those obtained by extending the arguments presented in \cref{scaling-nematics} to include friction, which would predict $\tilde E(q)\sim q^3$ at large scales $q\ll \lambda^{-1}$ (see \cref{external-fluid} below and \cref{Fig scalings-thin-layer}). Future work is required to clarify this point.

When friction becomes strong enough to screen the flow over distances comparable and even lower than the active length $\ell_{\text{a}}$, the number of defects decreases sharply (\cref{Fig defect-number}). Friction then stabilizes the flow into vortex lattices, and it organizes defects into dynamic yet positionally and orientationally ordered structures \cite{Doostmohammadi2016,Oza2016,Oza2016a,Thijssen2020b,Putzig2016,Patelli2019a,Srivastava2016a}. These structures are reminiscent of ordered phases of defects reported in microtubule suspensions experiments \cite{DeCamp2015,Pearce2020}. In this same context, other work has proposed to treat an active nematic as a fluid of defects \cite{Shankar2019}. Taking into account the active forces and torques between defects, Shankar and Marchetti developed a hydrodynamic theory of the defect fluid, and used it to predict transitions between a nematic, a chaotic, and a polar phase of defects \cite{Shankar2019}. Testing all these theoretical predictions and establishing the rich phase diagram of overdamped active nematics, including their defect phases, remains an experimental challenge for the future.

\subsubsection{Interaction with an external fluid} \label{external-fluid}

Beyond substrate friction, recent work has studied the effects of a different type of external dissipation on active nematic turbulence \cite{Martinez-Prat2021,Thijssen2021}. Instead of on a solid substrate, active nematic microtubule films can be assembled at an oil-water interface. Thus, the spontaneous flow in the active film induces flows in the adjacent passive fluids (oil and water) which, in turn, impact the flows in the active film via viscous forces. By varying the viscosity of the external oil, one can tune the external dissipation and vary the so-called viscous screening length $\ell_{\text{v}} = \eta/\eta_{\text{ext}}$, above which external dissipation dominates. Therefore, this setup provides a way to probe the wet-dry crossover in experiments.

As oil viscosity was increased, experiments found slower defect motion and smaller vortices \cite{Guillamat2016b}. In recent experiments where the oil has a free surface, measurements of the energy spectrum have also revealed a new scaling regime $\tilde E(q)\sim q$ at scales larger than the viscous length $\ell_{\text{v}}$ \cite{Martinez-Prat2021} (\cref{Fig microtubules}). This result has been explained by generalizing the theoretical framework of \cref{nematic-general,wet-nematics} to account for the viscous forces between the active film and the external fluids \cite{Martinez-Prat2021}. The generalized theory then fits the experimental spectra over a wide range of oil viscosities, and through scales across the wet-dry crossover \cite{Martinez-Prat2021} (\cref{Fig microtubules}). Besides explaining the measurements, the theory predicts up to six different scaling regimes, which are classified in terms of three lengths: the average vortex size $R_*$, the viscous length $\ell_{\text{v}}$, and the thickness $H$ of the external fluid layer (\cref{Fig scalings-thick-layer,Fig scalings-thin-layer}). When the external layer is thin, the scaling laws also depend on its boundary conditions; no-slip conditions then correspond to the case of friction with a solid substrate discussed in \cref{substrate-friction} (\cref{Fig scalings-thin-layer}). Finally, while three scaling regimes were observed in Ref. \cite{Martinez-Prat2021} (\cref{Fig microtubules}), three additional scaling laws await experimental verification.

\section{Conclusions and outlook} \label{conclusions}

Since its origins in bacterial suspensions, the field of active turbulence has thrived. Chaotic flows at low Reynolds number have been observed in a wide variety of active fluids (\cref{Fig1}). Accordingly, different classes of theories have been developed to understand the statistical properties of these turbulent-like flows. Here, we have reviewed all these scenarios, highlighting their similarities and differences, and their comparison with inertial turbulence. Although there is debate on the technical use of the term \emph{active turbulence}, here we have employed it in a loose sense to denote chaotic active flows. We note, however, that these flows represent a broad class of phenomena, which are described by a variety of model equations rather than being a single well-posed problem as for inertial turbulence.

We classified this variety of active turbulences according to whether systems have either polar or nematic order, and to whether they are described by models where momentum is either conserved (wet) or not conserved (dry). These key features define different scenarios of the transition from steady to turbulent flows: While it proceeds through non-oscillatory instabilities and defect proliferation in nematic fluids (\cref{transition-turbulence}), it generally involves oscillatory phenomena for polar fluids (\cref{active-polar-LC,active-polar-suspensions,Fig stability-polar-suspensions,Fig patterns-polar-suspensions}). This oscillatory behavior arises from active self-advection of the polar order. In fact, this advective nonlinearity of polar fluids is parallel to the velocity self-advection responsible for inertial turbulence. Consequently, active polar fluids at zero Reynolds number share advective mechanisms of energy transfer with passive fluids at high Reynolds number. In contrast, nematic fluids at zero Reynolds number lack self-advection terms. Thus, energy transfer in nematics is solely due to other nonlinear effects such as flow alignment and elastic stresses.

Besides instabilities and energy transfer, the differences between polar and nematic active turbulence are apparent in their scaling properties. Similar to Kolmogorov's scaling of inertial turbulence, active turbulence is also characterized by the emergence of power laws in the energy spectrum. However, in flocking-type models of polar fluids, these power laws have parameter-dependent exponents; their scaling properties are therefore non-universal (\cref{TTSH-scaling,Fig spectrum-TTSH}). This non-universality is due to the fact that, in addition to the self-advective nonlinearity, flocking-type models for polar fluid feature nonlinearities that allow for extra freedom in their spectra. In contrast, nematic fluids exhibit scaling laws with universal, parameter-independent exponents (\cref{scaling-nematics,external-fluid,Fig spectrum-nematic}). In this case, there are a few possible sets of exponents given by integer numbers (\cref{Fig scalings-thick-layer,Fig scalings-thin-layer}). Whether one or another set applies to a particular situation depends only on general properties of the system such as its dimensionality and the dominant mechanism of dissipation, including their wet or dry nature. A similar picture with universal exponents might apply to polar liquid crystals --- an issue to be addressed in future work. Altogether, in the bigger scheme of active matter physics, the scaling properties of active turbulence are a promising way of attaching simple numbers (exponents) to classes of active matter.

From this variety of scenarios, a generic picture of active turbulence emerges. In other forms of turbulence, including inertial and elastic, the fluid is driven externally by means of stirring, shaking, or shearing. Therefore, the spectrum of energy injection is externally imposed by the driving protocol. In contrast, active fluids are internally driven. Thus, the spectrum of energy injection is not imposed but results from a self-organized, autonomous process. Energy injection is therefore built into active turbulence equations in the form of terms that produce activity-driven instabilities. These instabilities feature characteristic length scales, which determine the primary scales of both energy injection and dissipation. Hence, active turbulence generally lacks energy cascades. The picture is strikingly different from that of inertial turbulence, where the scales of energy injection and dissipation are separated by a wide range of scales across which energy is transported. In this same range of scales, the energy spectrum behaves as a power law. In active turbulence, instead, injection and dissipation scales are not widely separated, and scaling regimes emerge not in between but on each side of the injection and dissipation scales.

Looking forward, this emerging picture still poses considerable challenges, for both theory and experiments. For example, future work is needed to clarify the regimes of applicability of the different theories. Perhaps the clearest example is that of bacterial suspensions, which have been described using both polar and nematic theories, and both in the wet and the dry limits. Other remaining tasks include searching for non-universal exponents in experiments, further characterizing the phase behavior and universal scaling regimes of active nematics, and beginning to do so in polar liquid crystals. Another important challenge across different systems is to experimentally measure elastic energy and energy transfer spectra, which also needs to be better understood theoretically.

Beyond these immediate challenges, we finish by outlining promising directions for future work. For example, the spectral properties of active turbulence in the time domain remain to be uncovered. In particular, the extent to which intermittency and anomalous exponents might be relevant in active turbulence is an open question. Along similar lines, analyzing active turbulence using metrics from chaos theory such as Lyapunov exponents and topological entropy \cite{Tan2019,Hashemi2021} might provide profound insights. Studying active turbulence on curved surfaces with different topologies, as in recent work \cite{Mickelin2018,Ellis2018}, might also reveal qualitatively new spectral features.

Along a different direction, active turbulence in 3D remains relatively unexplored. This situation, however, might change in the near future. For example, recent experiments on microtubule suspensions have revealed complex dynamics of topological disclination loops in 3D \cite{Duclos2020}, which might soon enable experimental studies of 3D turbulence. On the numerical side, simulations had been previously used to study the crossover between 2D and 3D nematic turbulence \cite{Shendruk2018}. Fully in 3D, more recent simulations have analyzed the formation of defect loops and their entanglement in dense defect networks \cite{Copar2019a,Binysh2020,Krajnik2020}. Understanding the connections between topological and flow structures in 3D and investigating their spectral properties are interesting avenues for future research.

Still in 3D, chirality generates active torques. As active forces, active torques and torque dipoles can induce flows and, for example, induce the rotation or translation of chiral active droplets \cite{carenza2019}. The effects of chirality on active turbulence have hardly been explored and might lead to surprising behaviors. Along these lines, recent work has shown that the spontaneous breaking of chiral symmetry leads to parity-violating Beltrami flows and an inverse energy cascade in 3D active fluids \cite{Slomka2017,Slomka2018}.

Finally, it is appealing to think about the role of active turbulence in biological systems \cite{Needleman2019}. On the one hand, chaotic flows might have to be tamed by biochemical regulation to enable robust biological functions and development \cite{Nishikawa2017a}. On the other hand, biological systems might also exploit active turbulence to enhance mixing and transport, thus overcoming the limitations imposed by flow reversibility in low-Reynolds hydrodynamics \cite{Groisman2001,C.P.2020}. When moving toward more complex biological systems, it will be interesting to understand how viscoelastic effects modify active turbulence \cite{Hemingway2015,Hemingway2016,Liu2021}, as well as to consider other forms of turbulence driven by active biochemical reactions in cells \cite{Tan2020a}.

\section*{Acknowledgments}

We apologize to the many colleagues whose work could not be cited owing to space constraints. R.A. acknowledges support from the Human Frontier Science Program (LT-000475/2018-C). R.A. acknowledges discussions in the virtual ``Active 20'' KITP program, supported in part by the National Science Foundation under grant No. NSF PHY-1748958. J.C. acknowledges support by MINECO (project PID2019-108842GB-C21) and Generalitat de Catalunya (project 2017-SGR-1061).

\bibliography{Turbulence}
\end{document}